\documentclass{aa}  

\usepackage{graphicx}
\usepackage{txfonts}
\usepackage{xcolor}
\usepackage{amsmath}
\usepackage{amssymb}
\usepackage{ulem} % Para tachar texto

\usepackage[colorlinks=true,citecolor=blue,urlcolor=blue]{hyperref}

\usepackage{txfonts}
\usepackage{booktabs}
\usepackage[utf8]{inputenc}

\begin{document}

   \title{Advanced classification of hot subdwarf binaries using artificial intelligence techniques and \textit{Gaia} DR3 data}
    \titlerunning{Classification of hot sds using AI}
    \authorrunning{Viscasillas V{\'a}zquez et al.}
   %\subtitle{}

\author{
C. Viscasillas V{\'a}zquez\inst{\ref{vilnius}} \and 
E. Solano\inst{\ref{cab}} \and 
A. Ulla\inst{\ref{uvigo},\ref{xeoma}} \and
M. Ambrosch\inst{\ref{vilnius}} \and
M. A. \'Alvarez\inst{\ref{cigus}} \and
M. Manteiga\inst{\ref{cigus2}} \and
L. Magrini\inst{\ref{oaa}} \and
R. Santove\~na-G\'omez\inst{\ref{cigus}} \and
C. Dafonte\inst{\ref{cigus}} \and
E. P\'erez-Fern\'andez\inst{\ref{uvigo}, \ref{beade}} \and
A. Aller\inst{\ref{oan}} \and
A. Drazdauskas\inst{\ref{vilnius}} \and
\v{S}. Mikolaitis\inst{\ref{vilnius}} \and
C. Rodrigo\inst{\ref{cab},$\dag$}.
}

\institute{
Institute of Theoretical Physics and Astronomy, Vilnius University, Sauletekio av. 3, 10257 Vilnius, Lithuania. \label{vilnius}
\and 
Centro de Astrobiolog\'ia (CSIC-INTA), Camino Bajo del Castillo s/n, E-28692 Villanueva de la Ca\~nada, Madrid, Spain. \label{cab}
\and
Applied Physics Department, Universidade de Vigo, Campus Lagoas-Marcosende, s/n, E-36310 Vigo, Spain. \label{uvigo} 
\and
Centro de Investigaci\'on Mari\~na, Universidade de Vigo, GEOMA, Edificio Olimpia Valencia, Campus Lagoas-Marcosende, E-36310 Vigo, Spain. \label{xeoma}
\and
CIGUS CITIC - Department of Computer Science and Information Technologies, University of A Coru\~na, s/n, E-15071 A Coru\~na, Spain. \label{cigus}
\and
CIGUS CITIC - Department of Nautical Sciences and Marine Engineering, University of A Coru\~na, Paseo de Ronda 51, E-15011 A Coru\~na, Spain. \label{cigus2}
\and 
INAF - Osservatorio Astrofisico di Arcetri, Largo E. Fermi 5, 50125, Firenze, Italy. \label{oaa}
\and
IES de Beade, Conseller\'ia de Educaci\'on e Ordenaci\'on Universitaria, Camino do Outeiro 10, E-36312 Vigo, Spain. \label{beade}
\and
Observatorio Astron\'omico Nacional (OAN), Alfonso XII 3, 28014, Madrid, Spain. \label{oan}
}

   \date{Received 25 June 2024 / Accepted 24 September 2024}

% \abstract{}{}{}{}{} 
% 5 {} token are mandatory
 
  \abstract
   %{Hot sds are compact blue evolved objects, burning He in their cores surrounded by a thiny H envelope. In the H-R Diagram they are located by the blue end of the Horizontal Branch. Most models agree on a quite probable common envelope binary evolution scenario in the Red Giant phase. However, the current binarity rate for this objects is a yet unsolved, but key, question in this field.}
    {Hot subdwarf stars 
    %(hot sds) 
    are compact blue evolved objects, burning helium in their cores surrounded by a tiny hydrogen envelope. In the Hertzsprung-Russell 
    %(H-R) 
    Diagram they are located by the blue end of the Horizontal Branch. Most models agree on a quite probable common envelope binary evolution scenario in the Red Giant phase. However, the current binarity rate for these objects is yet unsolved, but key, question in this field.}
   {This study aims to develop a novel classification method for identifying hot subdwarf binaries within large datasets using Artificial Intelligence techniques and data from the third \textit{Gaia} data release (GDR3). The results will be compared with those obtained previously using Virtual Observatory techniques on coincident samples.}
   %{The aim of this work is to develop a novel classification method for hot subdwarfs in binary systems, employing Artificial Intelligence techniques and comparing the results with those obtained previously on 
   %coincident samples using Virtual Observatory techniques.}  
   {The methods used for hot subdwarf binary classification include supervised and unsupervised machine learning techniques. Specifically, we have used Support Vector Machines (SVM) to classify 3084 hot subdwarf stars based on their color-magnitude properties. Among these, 2815 objects have \textit{Gaia} DR3 BP/RP spectra, which were classified using Self-Organizing Maps (SOM) and Convolutional Neural Networks (CNN). In order to ensure spectral quality, previously to SOM and CNN classification, our 2815 BP/RP set were pre-analyzed with two different approaches: 'cosine similarity' and Uniform Manifold Approximation and Projection (UMAP) techniques. Additional analysis onto a so-called ‘golden sample’ of 88 well-defined objects, is also presented.
   %Additionally, we have applied %other techniques such as 
   %a 'cosine similarity' technique to a so-called golden sample of 88 well-defined \textcolor{teal}{objects.}
   %hot sds stars.
   }
   {The findings demonstrate a high agreement level ($\sim$70-90\%) with the classifications from the Virtual Observatory Sed Analyzer (VOSA) tool. This shows that SVM, SOM, and CNN methods effectively classify sources with an accuracy comparable to human inspection or non-AI techniques. Notably, SVM in a radial basis function achieves 70.97\% reproducibility for binary targets using photometry, and CNN reaches 84.94\% for binary detection using spectroscopy. We also found that the single-binary differences are especially observable on the infrared flux in our \textit{Gaia} DR3 BP/BR spectra, at wavelengths larger than ${\sim}$700 nm.}
   {We have found that all the methods used are in fairly good agreement and are particularly effective to discern between single and binary systems. The agreement is also consistent with the results previously obtained with VOSA. In global terms, considering all quality metrics, the CNN is the method that provide the best accuracy. The methods also appear effective for detecting peculiarities in the spectra.  While promising, challenges in dealing with uncertain compositions highlight the need for caution, suggesting further research as needed to refine techniques and enhance automated classification reliability, particularly for large-scale surveys.}
  
  


   \maketitle
%
%-------------------------------------------------------------------
\section{Introduction}

Hot subdwarf stars (hot sds) are a unique stellar class characterized by their luminosity, which is lower than that of main-sequence stars of the same spectral type. \citet{Kuiper1939} and \citet{Humason1947} were the first to detect and catalogue these stars,
mainly classified into B- (sdB) and O- (sdO) types based on their atmospheric composition, with dominance of hydrogen (H) or helium (He), respectively \citep{Drilling13, Heber2016}.

%\textbf{Initially sdO/Bs have been found at high galactic latitudes \citep[e.g.][]{Bixler1991, Moehler1990}, but more comprehensive searches \citep[][and before]{Luo2021, Culpan2022, Dawson2024} have shown, that they are found in all Galactic populations. Recently, hot subdwarfs in GCs have been studied in quite some detail by \citet{Latour2023}.}

Initially found at high galactic latitudes \citep[e.g.][]{Bixler1991, Moehler1990}, more
comprehensive searches \citep[][and before]{Luo2021, Culpan2022, Dawson2024} have shown that
hot sds are found in all Galactic populations. Also, hot subdwarfs in globular clusters have been
studied in detail by \citet{Latour2023}. On the other hand, asteroseismology has provided relevant
insights into hot sds internal structure and evolution \citep[see][for a comprehensive
review of hot subdwarf pulsations]{Lynas-Gray2021}.

%The kinematics of hot sds was recently addressed by \citet{Bobylev2019} utilizing \textit{Gaia} Data Release 2 (GDR2) data based on the \citet{Geier19} sample. The authors categorized the data into low-latitude and high-latitude subsets, and found these to display kinematic characteristics of the thin and thick disks, respectively.
%Furthermore, hot sds play a significant role in the hot star population of old stellar systems like globular clusters \citep[e.g.][]{Remillard1979, Heber1987, Drukier1989, Bailyn1989} and elliptical galaxies \citep[e.g.][]{Dorman1995, Brown1997}. 

%On the other hand, asteroseismology of hot subdwarfs has provided relevant insights into their internal structure and evolution \citep[see e.g.][\textbf{for pulsating hot sdBs}]{Charpinet1996, Green2003, Oreiro2005, Ostensen2010}. \sout{In particular, three main types of pulsating hot B subdwarfs (sdBVs) have been identified: EC 14026 stars or p-mode short period pulsators, PG 1716-type or long period g-mode pulsators, and the hybrid sdBVs (with both p- and g-modes).}
%Additionally, a handful of the hotter variable sdO stars have been identified to show p-mode pulsations \citep{Dorsch2020}, \textbf{a phenomenon first sought in the works of \citet{RodriguezLopez2007}. \textbf{For a comprehensive review of hot subdwarf pulsations, see \citet{Lynas-Gray2021}.}}

In the Hertzsprung-Russell Diagram hot sds are located by the blue end of the Horizontal Branch \citep{Greenstein1974}, near the so-called Extended Horizontal Branch (EHB). With effective temperatures ($T_{\mathrm{eff}}$) 
%ranging from $\sim$20,000 to 40,000 K, 
exceeding $\sim$19,000 K, surface gravities (log$g$) in the range 4.5 $\leq$ log$g$ $\leq$ 6.5 dex, masses around 0.5 $M_{\odot}$ and a fraction of about 0.2 of the solar radius, they represent a late stage in stellar evolution, often formed when a red giant loses its outer hydrogen layers \citep{Heber2009,Heber2016}. As a consequence, these stars lack the capacity for sustaining hydrogen shell burning, and they deviate from the conventional evolution path, avoiding ascending the asymptotic giant branch (AGB) and proceeding directly onto the white dwarf (WD) cooling track \citep{Heber2016}. This suggests enhancement of mass-loss efficiency \citep{DCruz1996} or companion-driven mass transfer or coalescence. While \citet{Luo2024} conclude that the merging of double helium WD binaries can not explain the formation of C-deficient He-rich hot sds, most models agree on a quite probable common envelope binary evolution scenario in the Red Giant (RG) phase \citep[e.g.][]{Kramer2020}, because it is virtually impossible for a single RG to lose so much of its total mass by its own. \citet{Pelisoli2020}'s results suggest that the involvement of binary interaction is always necessary for the formation of hot sdBs. Indeed, decades of observations seem to support this assumption \citep[e.g.][]{Dworetsky1977,Paczynski1980, Ferguson1984, Kawka2015}. 
For sdOs, less common and hotter than sdB stars, and which are believed to have a carbon and oxygen core surrounded by a helium-burning shell, some examples have been found as central stars of planetary nebulae (CSPNe), even as binary systems \citep{Aller2013, Aller2015}. 

As the main objectives of our work deal with the classification of composite hot subdwarf systems, it
is worth indicating here some specific information on their formation and observational properties.
In particular, composite subdwarf B (sdB) + main-sequence (MS) systems are critical for understanding the formation and observational properties of sdB stars. The different scenarios of sdB/Os and binary evolution are described in detail in \citet{Han2002} and \citet{Han2003}. From population synthesis, a high binary fraction, up to 80\%, was predicted by these authors. More recent works on the helium-WD merger channel are e.g. \citet{Zhang2012}.

Companions to hot sds are of varied nature, from A-type stars to degenerate objects. These systems
typically form through binary interaction, as demonstrated in the works of colour \citet{Vos2012,Vos2013,Vos2017,Vos2018,Vos2020}, which characterized long-period sdB binaries and their evolution history. \citet{Pelisoli2020} further provided evidence that binary interaction is necessary for the formation of hot sds, highlighting the absence of wide sdB binaries that would suggest single-star formation scenarios. Additionally, recent findings by \citet{Lei2023} identified new long-period composite sdB binaries and emphasized the importance of binary interactions in their formation. Although inconclusive, the search for other types of companions to sdBs, including supermassive WDs (M > 1.0 M$_{\odot}$), neutron stars, black holes, brown dwarfs or even exoplanets, have been addressed by, for example, \citet{Silvotti2007,Geier2011,VanGrootel2021,Thuillier2022,Schaffenroth2022,Schaffenroth2023}.

%From the observational point of view on composite hot sds, a recent introduction can be found in
%\citet{Geier2022}. Early works, such as for example those by \citet{Thejll1995} and \citet{Ulla1998}, indicated that infrared flux excesses in the JHK bands were found to be mostly due to companion stars, mainly of spectral types A-K. An approximation to the distribution of hot sds according to their companions was given by \citet{Nemeth2020}, and space missions such as TESS \citep[Transiting Exoplanet Survey Satellite,][]{Ricker2015} are paving the way for the exploration of a
%larger number of variable –both binary and pulsating– hot subdwarf stars \citep{Sahoo2020,Schaffenroth2023,Uzundag2024}. However, the current binarity rate for this class of objects is yet unsolved, but key question in this field, and therefore, classification methods are essential.

From an observational perspective, composite hot sds have been extensively studied. Early works, such as for example those by \citet{Thejll1995} and \citet{Ulla1998}, indicated that infrared flux excesses in the JHK bands were found to be mostly due to companion stars, typically of spectral types A-K, for about 44\% of their sample. More recent studies \citep[see, e.g.][]{Nemeth2020}, indicate that hot sds in close binaries have either low mass K-M type or WD companions, while when in wide binaries, they have more massive F-G type companions. Typical binary periods of systems with low mass MS companions are less than 30 days, with the detection of reflection effects at the shortest periods –specially for M type companions. For the case of WD companions, ellipsoidal modulation and Doppler beaming have been found \citep{Schaffenroth2023}. On the other hand, wider hot sd binaries with orbital periods ranging from about 400 to 1500 days are the ones where more massive F-G type companions are found. In this case, double-lined composite spectra are often displayed and, in general, spectral decomposition techniques are required for proper analysis \citep{Nemeth2020}. When it comes to He content, radial velocity (RV) variability studies by \citet{Geier2022} revealed that He-poor hot sds display a high fraction of close binaries, while He-rich hot sds RV variability values are almost no significant, probably indicating different evolutionary channels. Among other striking results found, these authors indicate also the possible existence of a new binary subpopulation of hot sds cooler than about 24,000 K, with long orbital periods but late-type or compact companions. Further studies using space-based missions like the Transiting Exoplanet Survey Satellite \citep[TESS,][]{Ricker2015} have paved the way for discovering variable hot sds, both binary and pulsating, allowing for detailed analysis of their light curves, companion masses, and orbital parameters \citep{Sahoo2020, Schaffenroth2023, Uzundag2024}. These ongoing efforts are gradually improving our understanding of the binary nature and evolution of hot sds, highlighting the importance of refining classification methods.

The first catalogues including hot sds date back several decades \citep[see e.g.][]{Green1986, Kilkenny1988, Boyle1990, Kleinman2004, Mickaelian2008}. With the advent of the new spectroscopic surveys like the Massive Unseen Companions to Hot Faint Underluminous Stars from SDSS --Sloan Digital Sky Survey-- project \citep[MUCHFUSS,][]{Geier2011, Schaffenroth2018} or the Large Sky Area Multi-Object Fibre Spectroscopic Telescope \citep[LAMOST,][]{Cui2012, Luo2016}, and space missions like \textit{Gaia} \citep{Geier19, Culpan2022}, or combinations of them \citep{Lei2018}, their census has increased significantly. This means that methodologies to analyse 
%the 
photometry and spectra must adapt to the large amounts of data provided \citep[see e.g.][]{Ambrosch2023}.

Simultaneously with the exponential growth of data, initiatives such as the Virtual Observatory\footnote{\url{http://www.ivoa.net}} (VO) have proven to be very useful in facing new frontiers in the field of massive data analysis. 
%In a study led by \citet{Oreiro2011} and using the VO methodology, a custom procedure was designed\citep[VO,][]{Rodrigo2017}
\citet{Oreiro2011}, using the VO methodology, designed a custom procedure to discover previously uncatalogued hot sds within blue object samples. This approach minimized contamination on WDs, cataclysmic variables, and OB stars. 
%Expanding this base, 
\citet{Perez-Fernandez2016} further refined the methodology and successfully identified new hot sds with SDSS spectra. Next, \citet{Solano22} presented a method for identifying binary systems involving hot sds, utilizing the VOSA Virtual Observatory tool \citep{Bayo2008}. The approach involves constructing the Spectral Energy Distribution (SED) from ultraviolet to infrared wavelengths, 
%\citep{Rodrigo2014, Rodrigo2020}, 
identifying binaries through flux excess towards redder bands, and estimating physical parameters for both individual and composite hot sds based on the optimal fitting model \citep{Rodrigo2020}.

Recent studies have leveraged artificial intelligence (AI) techniques for the identification and classification of hot subdwarf stars, significantly enhancing automation and accuracy. \citet{Bu2019} employed a method combining convolutional neural networks (CNN) and support vector machines (SVM) to analyse LAMOST DR4 spectra, achieving an F1 score of 76.98\% and outperforming other machine learning algorithms. Similarly, \citet{Tan2022} developed a robust identification method using a hybrid CNN model on LAMOST DR7-V1 data, attaining an accuracy of 87.42\% in identifying new candidates. These approaches highlight the effectiveness of AI in the large-scale spectral classification and discovery of hot subdwarf stars.
As a continuation, we conduct an in-depth analysis of binary-single hot sds classification utilizing the datasets provided by \citet{Solano22} addressed with machine learning techniques, which are becoming indispensable in many branches of astronomy. Our analysis incorporates the use of the recently released \textit{Gaia} DR3 data, with a particular focus on the processing of BP/RP (red/blue photometer) spectra through state-of-the-art machine learning (ML) and AI techniques. This is the first instance of using \textit{Gaia} BP/RP spectra for this purpose, leveraging the most comprehensive dataset to date.

Specifically, we have used Support Vector Machines (SVMs) to classify 3084 hot sd stars based on their colour-magnitude properties, as well as Self-Organizing Maps (SOMs) and Convolutional Neural Networks (CNNs) to classify GDR3 BP/RP spectra for a subsample of 2815 objects for which spectra were available.
CNNs and SVMs have emerged as a powerful tool in astronomy \citep[see e.g.][for a review]{Ball2010}, showcasing its increasing potential in recent years of which our study could be taken as example.
Our methodology also includes the \textit{Gaia} Utility for the Analysis of self-organizing maps (GUASOM) \citep{Fustes14, Alvarez22} specially designed for the treatment and analysis of massive \textit{Gaia} data.  
Additionally, we have applied a 'cosine similarity' and Uniform Manifold Approximation and Projection (UMAP) techniques to scrutinize the BP/RP spectra and a so-called 'golden sample' of 88 well-defined, both single and binary, hot subdwarf stars. \par
Given the above, addressing the different classification methods employed, this paper is structured as follows:
photometry, tackled with SVMs, goes in section $\ref{sec:photometry}$ while spectroscopy goes in section $\ref{sec:spectroscopy}$, with subsections $\ref{subsection:Pre-analysis of the spectra}$, $\ref{subsec:som}$ and $\ref{sec:CNN}$ for spectra pre-analysis and for the SOM and CNNs techniques, respectively. In section $\ref{sec:comparison}$ we compare all methods with the results previously obtained with VOSA, using different prediction metrics to evaluate the performance of our classification models. In section $\ref{sec:88}$ we apply the 'cosine similarity' method to our smaller and well-defined golden subsample. Finally, in section $\ref{sec:conclusions}$, we summarize the major outcome of this work and give the conclusions.

%--------------------------------------------------------------------

\section{Single-binary classification of hot sds based on \textit{Gaia} BP/RP photometry using ML techniques}
\label{sec:photometry}
Machine learning has yielded several algorithms proficient in effectively handling large datasets for classification tasks. Some common classification algorithms include random forest, decision trees, k-nearest neighbours (k-NN), and Naive Bayes. One of them, called support vector machines (SVMs) and developed at Bell Laboratories for other purposes \citep{cortes1995}, has proven to be also a very useful tool for classification tasks in astronomy \citep[see e.g.][for a review]{Zhang_2014}. 
Some examples include its use in classification of stars, galaxies, quasars, AGNs, and gravitational lenses \citep[see e.g.][]{Huertas2008, Huertas2011, Hartley2017, Marton2019, Wang2022, Viscasillas23, Hassanshahi2023}. After testing several algorithms, we evaluated their accuracy, finding that the SVMs yielded the highest validation accuracy for our case. Given that model accuracy is our primary concern, we opted for the SVM algorithm for the photometry case. In essence, the SVM algorithm constructs a hyperplane in the feature space to maximize a margin, effectively partitioning data points into separate regions. The margin refers to the distance between the separating hyperplane and the closest data points from each class, also known as support vectors. The main objective of SVM is then to maximize that margin by finding the optimal hyperplane that best separates the classes in the feature space. This partitioning enables the algorithm to assign labels to each region, and it subsequently assesses label accuracy by comparing them against actual samples and computing the loss function \citep[see e.g.][for a detailed explanation and further reading]{boser92,cristianini2000}.

\subsection{The single-binary classification boundary based on the colour-magnitude diagram}
\label{subsec:color-mag}
\begin{figure}
  \resizebox{\hsize}{!}{\includegraphics{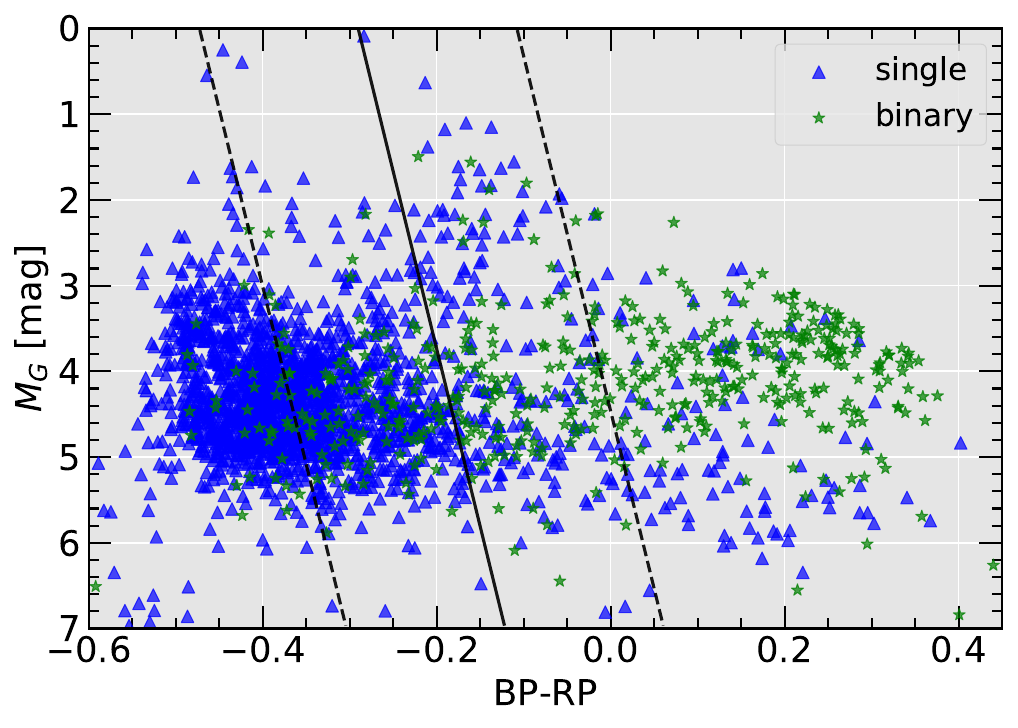}}
  \caption{Colour-magnitude diagram showing the 3084 objects included in \citet{Solano22}. Single stars are plotted as blue triangles while binaries are plotted as green stars. The central line is the optimal hyperplane that best separates both populations, computed using support vector machines (SVMs) with a linear kernel. The dashed lines are the positive and negative bounding hyperplanes that separate the soft margin.}
  \label{fig:SVM_linear}
\end{figure}

\begin{figure}
  \resizebox{\hsize}{!}{\includegraphics{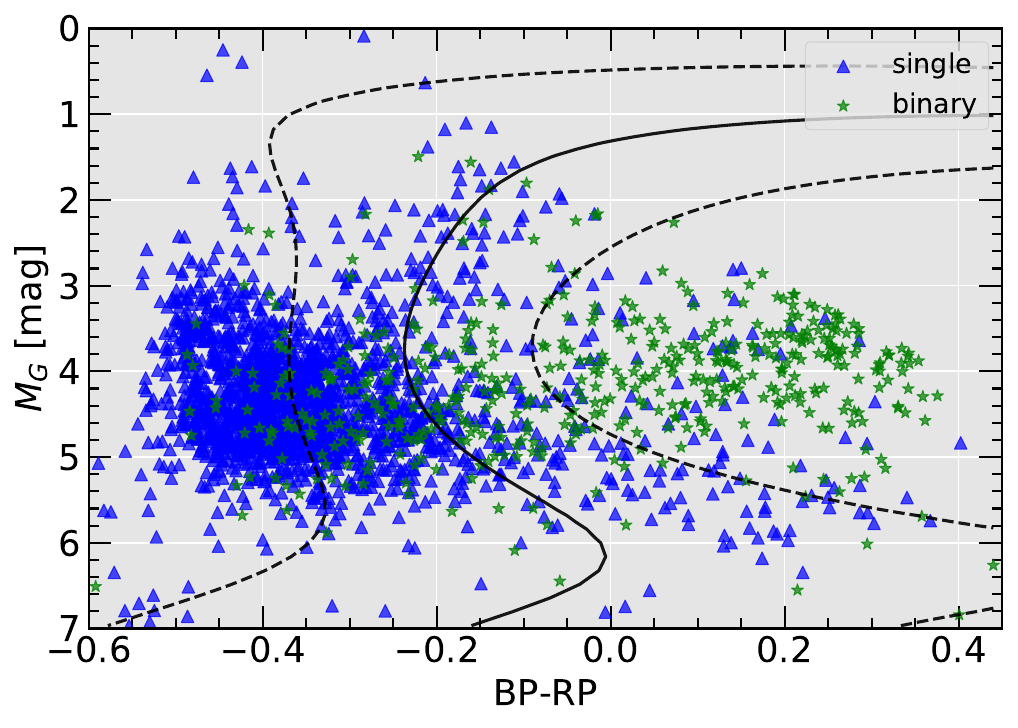}}
  \caption{Same as Fig. \ref{fig:SVM_linear} but using a radial basis function (RBF).}
  %Color-magnitude diagram showing the 3084 \citet{Solano22} single-binary objects. The curve is the hyperplane that best separates both populations computed using support vector machines (SVMs) with a radial basis function (RBF). Colours and symbols as in Fig. \ref{fig:SVM_linear}}
  \label{fig:SVM_radial}
\end{figure}

\citet{Geier19} proposed a single-binary hot sds frontier based on a colour-magnitude diagram (CMD) where, mainly, single stars are located at $G_{BP}$ $-$ $G_{RP}$ $\leq$ 0.0 and the binaries with cooler companions are found in the redder region. We are now trying to optimize that margin with the help of ML. Starting from the sample of \citet{Solano22}, composed of 3084 hot sds (2469 singles and 615 binaries classified using VOSA), we computed a new single-binary separation and membership probability to each class using the aforementioned SVMs analysis. This allowed us to find the optimal hyperplane that best captures the training data and separates the stars into singles and binaries, maximizing the margin between the two classes. Since the separation of both classes is not perfect and singles and binaries appear mixed, we use a soft margin, which allows some points to be misclassified. This is done by including a penalty parameter to control the tolerance of the classification, and allowing outliers to exist in the opponent classification. Indeed, and as shown in the Figs. \ref{fig:SVM_linear} and \ref{fig:SVM_radial}, there is some overlap in the colour-magnitude diagram proposed by \citet{Geier19} between objects classified as singles and binaries (this is specially visible in the central part of the CMD). On the other hand, in datasets with skewed class distributions imbalanced like ours (single/binary ratio of approx. 80/20\%), the margin tends to favour the majority class, in our case singles. For this reason, we used a weighted (cost-sensitive) SVM, which improved the classification results. We trained the SVM using different SVM Kernels \citep{Scholkopf1997} and implemented it using the {\sc scikit-learn} package \citep{scikit-learn11}. We found that the accuracy of the models compared with that of VOSA is better when we use a linear kernel (Fig.~\ref{fig:SVM_linear}) and a Gaussian radial basis function (RBF) (Fig. \ref{fig:SVM_radial}) than when we use a 3-degree polynomial and a sigmoid-shaped curve kernel. In Section $\ref{sec:comparison}$ we quantify the classification results using various prediction and quality metrics.

In Eq. $\ref{eq:decision_function}$ we show the decision function f(x) for the weighted-SVM (w-SVM) linear case, which determines the membership of a star to one of the classes in the feature space. It follows the general form f(x) = w$^\mathrm{T}$ $\cdot$ x + b, where w is the vector of coefficients that weights the features, x$_{1}$ and x$_{2}$ the colour BP-RP and absolute magnitude M$_{G}$ respectively, and b is the bias term that adjusts the location of the decision function relative to the origin. This decision function represents a linear separation between classes single-binary, where the signs of the resulting values of f(x) indicate which class the star belongs to: if f(x) > 0, the point is classified 
%into 
as binary, and if f(x) < 0, it is classified 
%into 
as single.

\begin{equation}\label{eq:decision_function}
f(x) = 5.48 \cdot x_1 - 0.13 \cdot x_2 + 1.59
\end{equation}

Due to the nonlinear transformation induced by the RBF kernel, expressing the decision boundary as a simple linear equation is not feasible.

\subsection{Extrapolation to a larger catalog}
\begin{figure}
  \resizebox{\hsize}{!}{\includegraphics{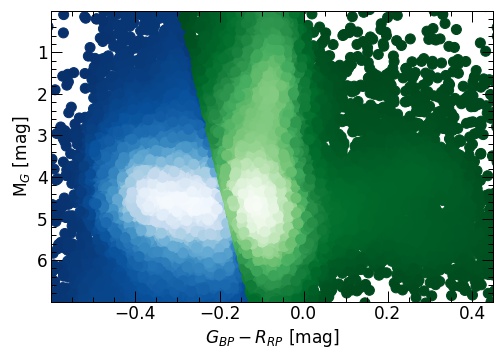}}
  \caption{Colour-magnitude diagram with the sample of 39,800 hot subluminous star candidates selected in \textit{Gaia} DR2 by \citet{Geier19} using a linear w-SVM classification. The stars in blue colour correspond to a prediction of singles and the green ones to binaries. The data is coloured using a point density function.}
  \label{fig:P_linear}
\end{figure}

\begin{figure}
  \resizebox{\hsize}{!}{\includegraphics{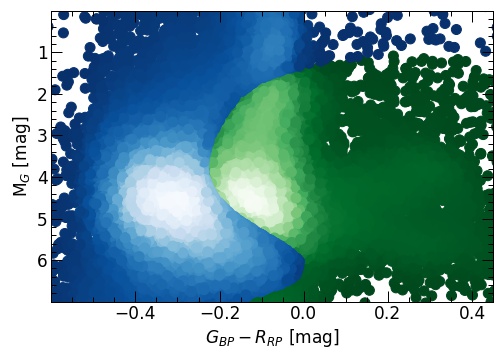}}
  \caption{Colour-magnitude diagram with the sample of 39,800 hot subluminous star candidates selected in \textit{Gaia} DR2 by \citet{Geier19} using a radial basis function (RBF) for the w-SVM classification. Colours as in Fig. \ref{fig:P_linear}.}
  \label{fig:P_radial}
\end{figure}

We applied the two previous methods (w-SVM RBF and w-SVM linear), trained with the aforementioned sample of 3084 hot sds of \citet{Solano22}, to a larger catalogue which contains $\sim$39,800 hot subluminous star candidates selected in \textit{Gaia} DR2 \citep[][see Figs.~\ref{fig:P_linear} and \ref{fig:P_radial}]{Geier19}. When we use the w-SVM linear we get a single/binary ratio of 51/49 while if we use the w-SVM RBF we get 70/30.

%\sout{Then we calculated the single-binary membership probabilities calibrated using \citet{PlattProbabilisticOutputs1999} scaling extended for multi-class classification \citep{Wu04} and whose results are shown in Figs. \ref{fig:P_linear} and \ref{fig:P_radial}.} \textcolor{teal}{\sout{However, it is important to note that these probabilities do not account for the weighted (cost-sensitive) SVM approach.}} 

Since we are testing candidates, all of the above proves to be effective for a first and raw classification and statistical separation of large amounts of stars into possible single-binaries, for a subsequent detailed analysis of the spectra using other techniques, as we will see below. Even so, the prediction accuracy may decrease when extrapolating beyond the feature space region covered by the training samples \citep{Wang2022}. Combining photometry and spectroscopy for star classification offers advantages over only using spectra. It reduces initial complexity by employing less computationally intensive photometric data for a preliminary classification and enhances precision by leveraging complementary information from both data types, providing flexibility for optimizing each stage independently.

\section{Single-binary classification of hot sds based on \textit{Gaia} DR3 BP/RP spectroscopy using SOM and CNN}
\label{sec:spectroscopy}
In the previous section, our investigation primarily delved into photometry. However, the transition to spectroscopy with \textit{Gaia}'s BR/RP spectra was prompted by the need for more detailed insights into stellar characteristics. Spectroscopic data offer richer information, which are crucial for distinguishing between single and binary systems. Notably, the use of \textit{Gaia}'s spectra for this purpose is novel, as previous studies have predominantly relied on photometric observations.  Therefore, our exploration  utilizing \textit{Gaia}'s spectra for discriminating between single and binary stars fills a notable gap in the field, offering a promising avenue for enhanced stellar characterization. In the next sections, we address this by employing two techniques, Self-Organizing Maps (SOM) and Convolutional Neural Networks (CNN), to effectively analyse \textit{Gaia}'s BR/RP spectra. To train the SOM and CNN, we used 2815 objects of the \citet{geier20} catalogue having binary/single classification in \citet{Solano22} and with \textit{Gaia} DR3 BP/RP spectra \citep{Angeli23}. The \textit{Gaia} DR3 BP and RP spectra span the wavelength ranges of 330-680 nm and 640-1050 nm, respectively, with resolutions ranging from 100 to 30 for BP and from 100 to 70 for RP \citep{Angeli23}. To normalize these spectra, we divided each flux value by the maximum flux value within that spectrum. By doing this, we ensure that the highest flux value in each of the spectra is scaled to 1, while all other flux values are proportionally reduced, allowing a consistent comparison between different spectra. \par
Both the SOM and CNN techniques are purely data-driven. This means that their classification is only based on the numeric flux values in the spectra, with no additional input physics or other information about the nature of the spectra involved. 
 
\subsection{Pre-analysis of the spectra} 
\label{subsection:Pre-analysis of the spectra}
To ensure that SOM and CNN will be trained on good quality data, initially we pre-analysed our BP/RP spectra with two different approaches: A Uniform Manifold Approximation and Projection (UMAP) method and the 'cosine similarity' analysis. \par
\begin{figure}
    \centering
    \includegraphics{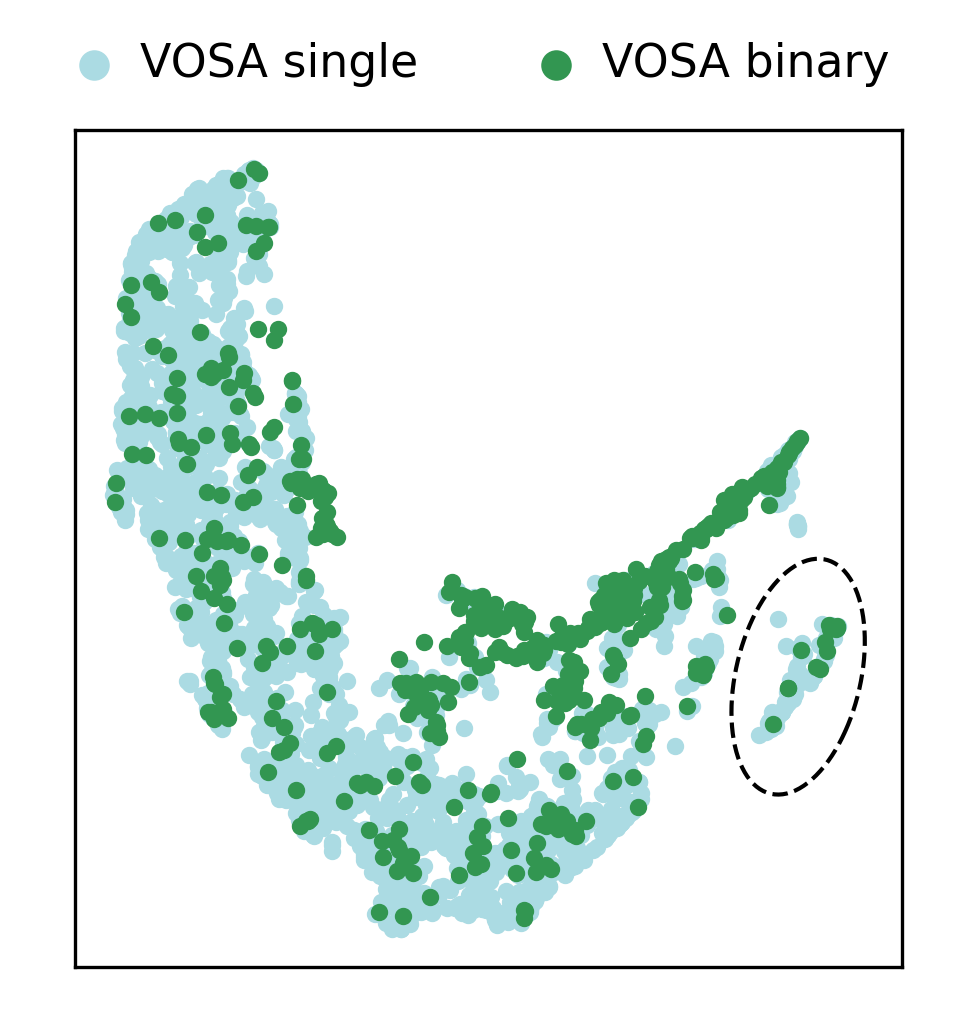}
    \caption{UMAP projection of our 2815 sample spectra. Every data point in the map represents a spectrum. Colours indicate if the spectrum has been classified as single or binary by VOSA. The dashed ellipse marks the position of the 98 anomalous spectra in the projection. The axis dimensions have no direct physical meaning.}
    \label{figure:UMAP 2815}
\end{figure}

The UMAP tool is a method to visualize similarities between high dimensional data points in large data sets \citep{mcinnes2018}. In our context, every spectrum is a 308 dimensional data point (one dimension for every wavelength bin in a spectrum) in a data set of size 2815. In a UMAP projection, each of the high-dimensional data points is represented by a single point in a two-dimensional plane. The distance between two points in this plane is a measure of the similarity of the represented spectra. Close points in the UMAP projection represent similar spectra, while a large distance between points shows that the spectra are different from each other. Figure~\ref{figure:UMAP 2815} shows the UMAP projection of our 2815 sample spectra. We can see that there is a variety of spectral shapes in our data set, with an  
%However, there are no single outlier spectra that are very distant from the rest of the points. 
isolated group of 98 spectra. Our investigation of this group shows that all of them show atypically low flux values at wavelengths <~400~nm. These low flux values emerge during \textit{Gaia}'s data reduction process and are not of physical origin \citep[see e.g.][]{vanLeeuwen22, Angeli23}. We will show in Sect.~\ref{sec:Relative difference between singles and binaries} that the difference between single and binary spectra increases towards larger wavelengths, and that they are indistinguishable at wavelengths <~400~nm. We concluded that for the single-binary classification, the spectral flux at these low wavelengths is not significant. Therefore, we decided to keep these 98 anomalous spectra for the training of our SOM and CNN. We also see in Fig.~\ref{figure:UMAP 2815} that single and binary spectra cover the same space in the UMAP projection. There is a concentration of binary spectra along a filament in the projection, but this filament also contains a significant number of single spectra. Binary spectra are also found outside of the mentioned filament, distributed more or less uniformly among the single spectra. This UMAP approach is therefore not able to reliably classify individual, unlabelled spectra into single or binary. \par
The cosine similarity measure is a different method to assess the similarity between data points in high-dimensional sets. We implemented the method with the {\sc scikit-learn} package \citep{scikit-learn11}. To do this, we considered the 2815 normalized spectra as vectors and derived the 'cosine similarity' by using the following Euclidean dot product formula (Eq. \ref{eq:cosine_similarity}):

\begin{equation}\label{eq:cosine_similarity}
S_{C}(\mathbf{A}, \mathbf{B}) = \cos(\theta) = \frac{\sum_{i=1}^{n} A_{i}B_{i}}{\sqrt{\sum_{i=1}^{n} A_{i}^2} \cdot \sqrt{\sum_{i=1}^{n} B_{i}^2}}
\end{equation}

where $A_{i}$ and $B_{i}$ are the $i$-th components of vectors $\mathbf{A}$ and $\mathbf{B}$, respectively. For every pair of spectra in our data set, this method returns a value between 0 (very dissimilar) and 1 (identical). We can average the cosine similarity values between a given spectrum and all other spectra, to find how similar this single spectrum is to the whole data set. Outlier spectra will then have a low mean cosine similarity value. The vast majority of spectra (96\%) have similarity values >~0.97, meaning that they are very similar to each other. Only a few spectra are dissimilar to the rest.

\begin{figure}
    \centering
    \includegraphics{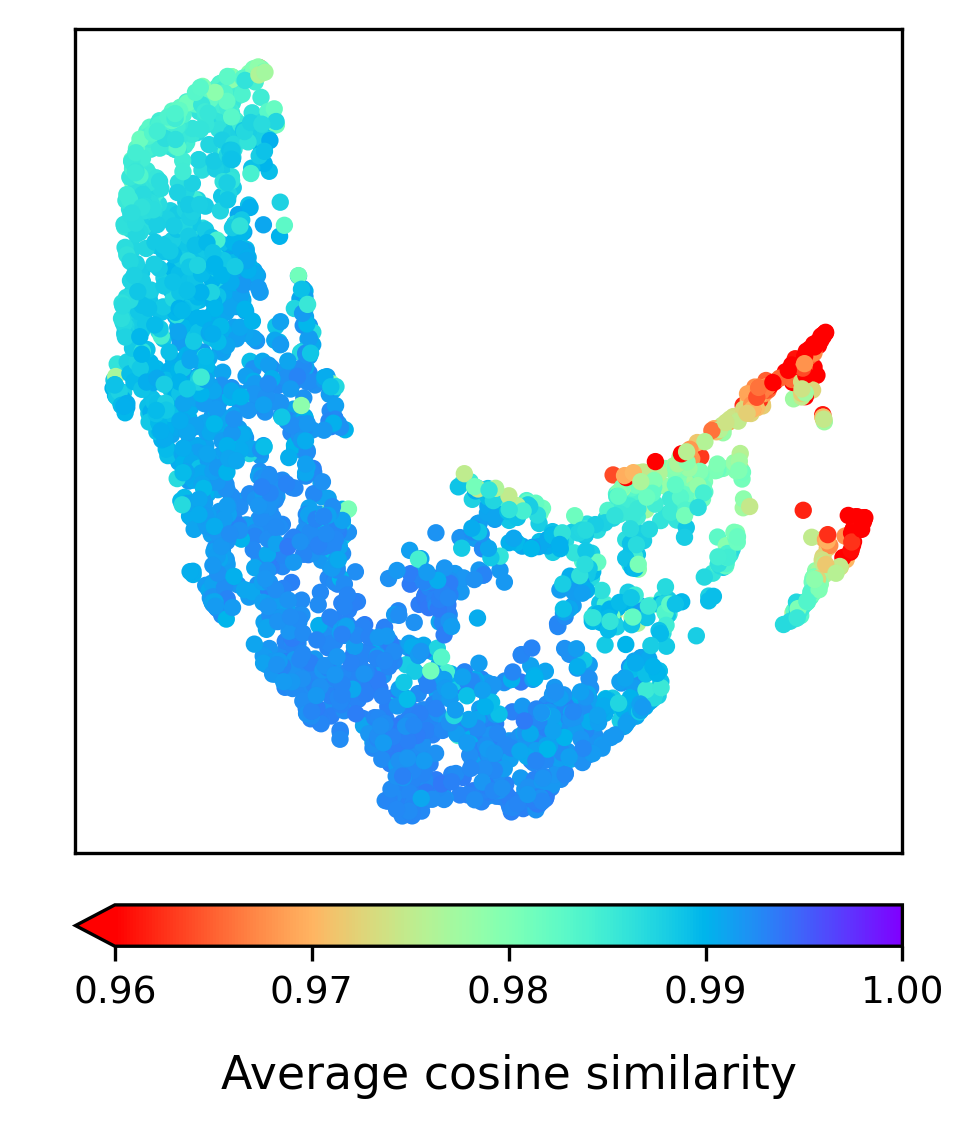}
    \caption{Same UMAP projection as in Fig.~\ref{figure:UMAP 2815}. Colour-coding indicates the mean cosine similarity of every projected spectrum.}
    \label{figure:cos-umap}
\end{figure}

We compare the results from our cosine similarity analysis and the UMAP projection in Fig.~\ref{figure:cos-umap}. Here, we colour-code the data points in the UMAP by the mean cosine similarity for every spectrum. The colour-coding shows that the two methods agree well with each other: Spectra at the edges of the UMAP also have lower values of the mean cosine similarity. The 98 anomalous spectra, identified in the UMAP, also have low cosine similarity values. The average cosine similarity of the group of anomalous spectra is 0.971, while the average similarity of the rest of the spectra is 0.988. This comparison shows that the cosine similarity and UMAP analysis complement each other and are effective in finding outliers in a large set of spectra. \par

For future projects, we aim to apply SOM and CNN to sets of unlabelled spectra that are much larger than our current set of 2815. Then it will be impossible to visually inspect all spectra prior to the training. The cosine similarity and UMAP analysis will therefore be valuable tools to assess the quality of our initial data set.

\subsection{SOM classification based on the GDR3 BP/RP spectra}
\label{subsec:som}
With the aim of classifying stars based on their nature, without any other a priori knowledge that could bias the results, we have used an unsupervised learning technique called Self-Organizing Maps (SOM).
SOM is a technique that unifies the concepts of clustering and dimensionality reduction, since it groups objects based on their similarity and, in turn, performs a non-linear reduction of dimensionality by projecting the data into a certain number of groups, called neurons, and retaining the information on the distribution of the data in their topology.
Each neuron has a prototype, that is the representative pattern of the objects that belong to that group, and normally the neurons are arranged in a two-dimensional structure,

The optimal configuration of the SOM, presented in this paper, is found to be around 100 clusters in a 10x10 lattice, and it has been trained for 200 iterations.
%\textcolor{magenta}{ESM: ¿cluster aquí es lo mismo que neurona? Si es así yo utilizaría neuron por homogeneidad}

We train the SOM on our sample of 2815 \textit{Gaia} DR3 BP/RP spectra together with their classification from the \citet{geier20} catalogue. We then analyse and visualize the SOM outputs with the \textit{Gaia} Utility for the Analysis of self-organizing maps (GUASOM) \citep{Fustes14,Alvarez22}.

\begin{figure}
    \centering
    \includegraphics[width=0.9\columnwidth]{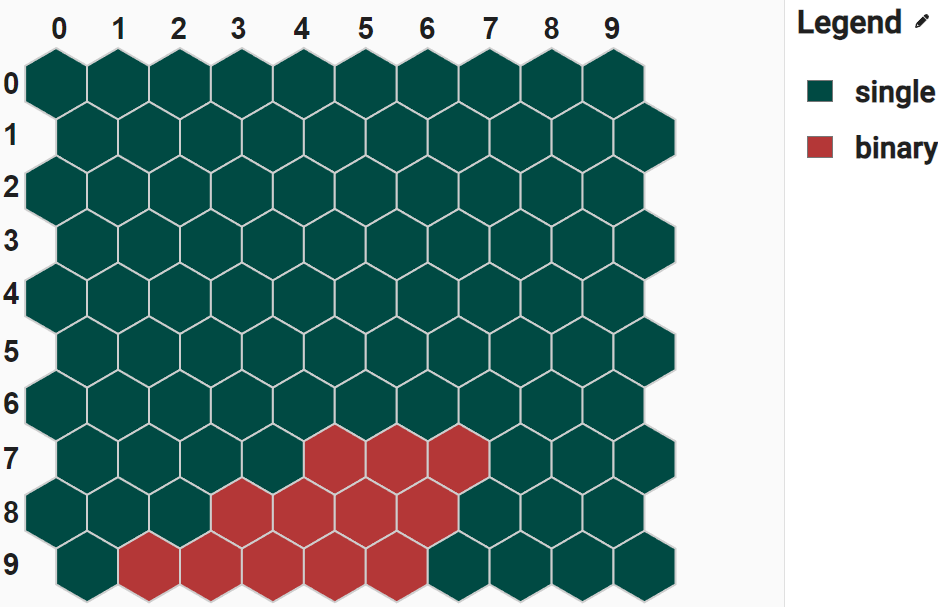}
    \caption{Category plot, showing the representative label of each neuron according to classification given in \citet{Solano22}.}
    \label{figure:som_catalogue_plot}
\end{figure}

Figure~\ref{figure:som_catalogue_plot} presents a visualization of the results obtained showing how the technique has been able to satisfactorily group the different types of stars using their spectra. Furthermore, by labelling the different neurons we can easily identify the region of neurons that mostly represents the binaries and the region of the single stars.

\begin{figure}
    \centering
    \includegraphics[width=0.9\columnwidth]{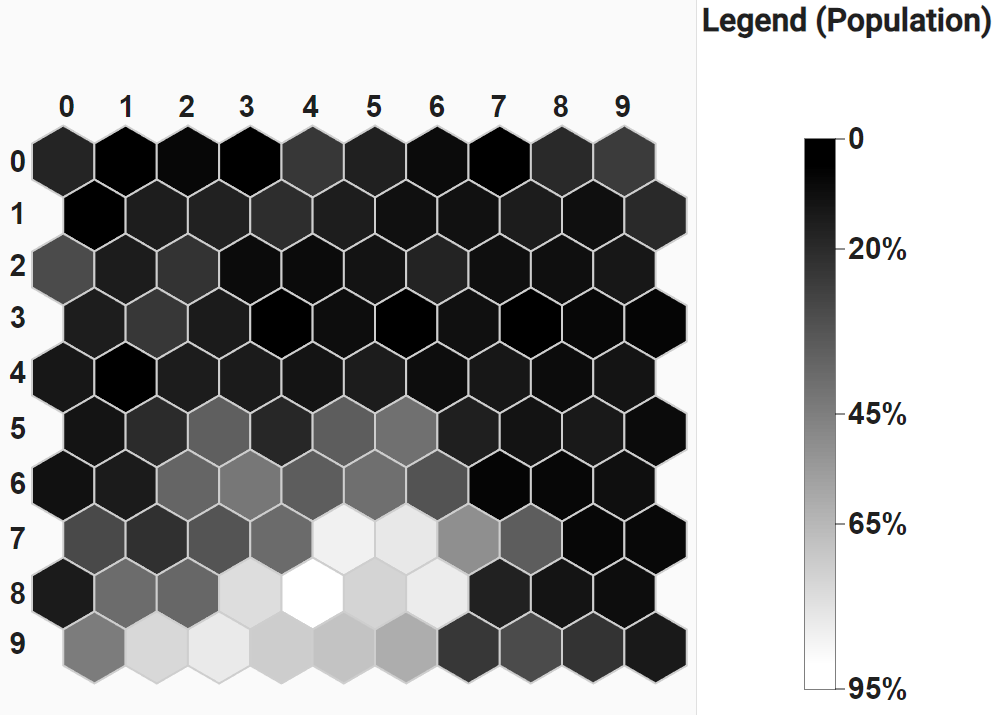}
    \caption{Binary probability distribution: it shows the probability of each neuron to represent a binary star, where whiter colours represent higher probabilities.}
    \label{figure:som_binary_probability}
\end{figure}

%\textcolor{magenta}{Remove the "unknown" label from the legend}
Labelling is carried out by absolute majority, therefore, to analyse the probabilities that each neuron has of representing a binary or a single star, Fig.~\ref{figure:som_binary_probability} should be analysed.

%\textcolor{magenta}{How have the probabilities in Fig8 been calculated?}

\begin{figure}
    \centering
    \includegraphics[width=0.8\columnwidth]{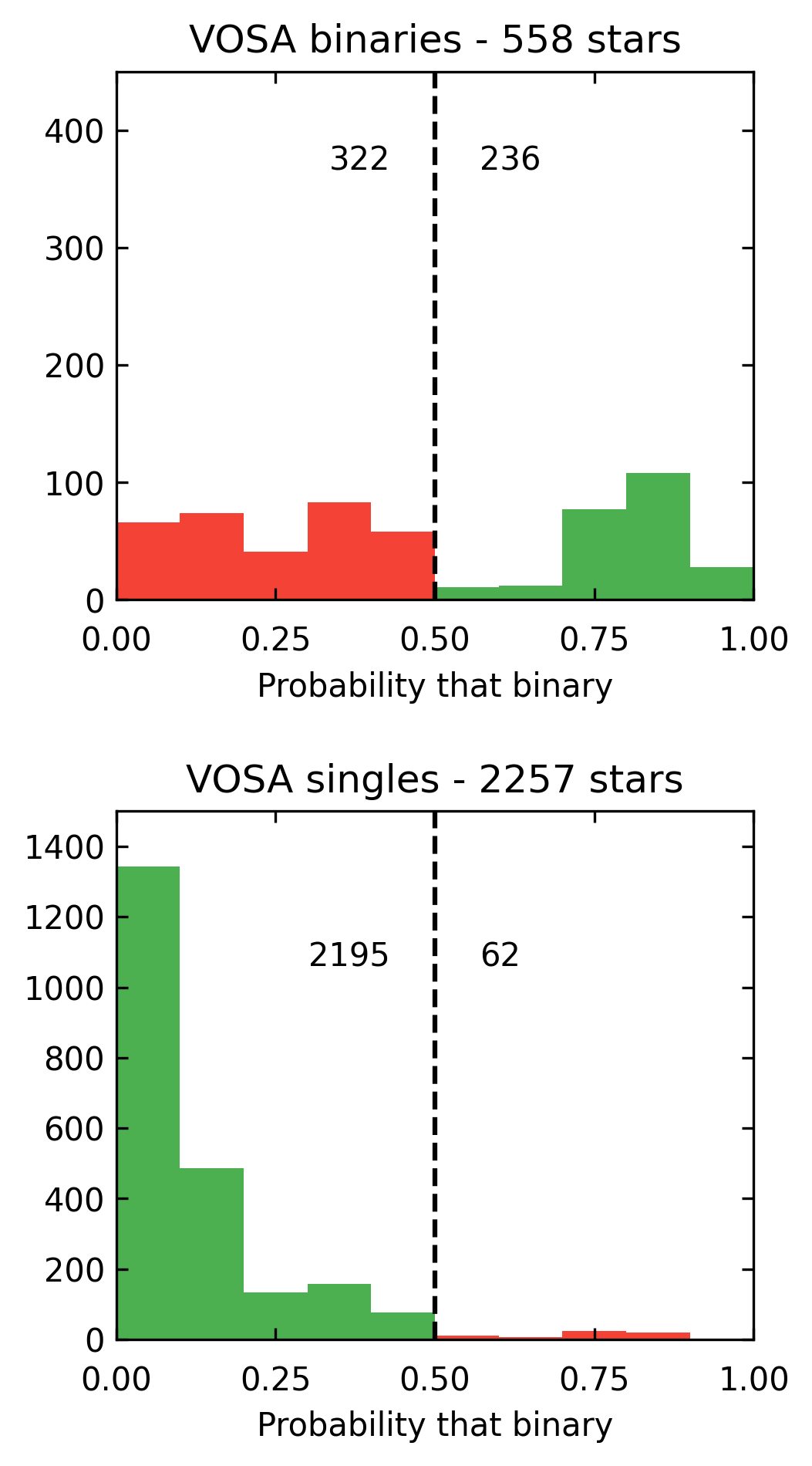}
    \caption{Distribution of SOM output probabilities for spectra to be "binary". Top panel: Distribution of probabilities for spectra that have been labelled binary by VOSA. Bottom panel: Same as top panel, but for VOSA single spectra. Numbers in the plot show how many spectra fall above or below the 50\% probability threshold. Green bars highlight agreement between SOM and VOSA labels, red bars show disagreement between the two methods.}
    \label{figure:netork_predictions_som}
\end{figure}

Figure~\ref{figure:netork_predictions_som} shows that our SOM method is able to identify single stars in our data set with high accuracy. More than 97\% of the VOSA single stars have been labelled correctly. This in turn means that the number of VOSA single stars that have wrongly been labelled binary by SOM is only 3\%. We provide a more in-depth analysis of the prediction metrics in Sect.~\ref{sec:comparison}.

\subsection{
%Convolutional neural network (CNN)  
CNN classification based on the GDR3 BP/RP spectra}
\label{sec:CNN}
%\textcolor{red}{Markus}

Convolutional neural networks have been successfully used to classify objects in large astronomical data sets. Examples are the morphological classification of galaxies based on 2D images \citep{Khalifa2018} and the prediction of stellar spectral and luminosity classes from 1D spectra \citep{Sharma2019}. 
\cite{Tan2022} use a CNN to identify hot sds in a large sample of spectra of mixed spectral types from the LAMOST survey. We further specialize the CNN approach to classify subdwarf spectra into single and binary. \par
The convolution layers enable a CNN to identify extended features in input data. The network can then learn the connections between these features and the class of the object which is represented by the data. With network gradients it is possible to visualize the importance of different features in the input data on the network output class. \par
We trained a CNN to classify our sample of 2815 GDR3~BP/RP spectra into single and binary. Neural networks are a class of supervised machine learning techniques. Supervised techniques require a training set with pre-determined labels, from which the network can learn. In our case, this training set consists of GDR3~BP/RP spectra and their associated labels (these are "single" or "binary"). Once the training is finished, we can use the trained network to classify new, previously unlabelled spectra. \par
Our CNN behaves like a function, that returns a value between 0 and 1 for every input spectrum. This value can be interpreted as a membership probability to the class "binary". \par
For the training of our network, we used a set of 700 spectra and their VOSA labels. 
To avoid the complications that arise from training on imbalanced training data (see, for example \citealt{Krawczyk2016}), our training set contains as many binary stars as single stars. The number of available training spectra was therefore restricted by the relatively low number of spectra labelled as "binary" in our overall sample. \par
To monitor the learning progress of the network during the training, and to detect possible overfitting to the training set, we also constructed a validation set of 300 spectra and labels. This validation set is also balanced with respect to single and binary samples. Spectra from our overall sample were assigned at random to either the training set or validation set. 
We tested different network architectures and hyperparameters to optimize the performance of our CNN. The final architecture and the chosen hyperparameters are listed in Tab.~\ref{table:CNN_architecture_hyperparameters}.

\begin{table}[]

\caption{Architecture and most important hyperparameters of our CNN.}
\centering
\begin{tabular}{@{}ll@{}}
\hline
\hline
\noalign{\smallskip}
Layer          & Hyperparameters                                                                                 \\ 
\hline
\noalign{\smallskip}
1D convolution layer & \begin{tabular}[c]{@{}l@{}}filters=16,  kernel\_size=20,\\ activation="LeakyReLU"\end{tabular} \\
\noalign{\smallskip}
flatten              &                                                                                                \\
\noalign{\smallskip}
Dense layer          & units=8, activation="LeakyReLU"                                                                \\
\noalign{\smallskip}
Dense layer (output) & units=1, activation="sigmoid"                                                                  \\ 
\hline

\end{tabular}
\label{table:CNN_architecture_hyperparameters}
\end{table}

The convolution layer is designed to identify features in the input spectra, such as absorption lines and positive or negative slopes. Based on these found features, the following dense layers calculate the output probability that a spectrum is "binary". The optimal pixel values of the convolution filters and the weights and biases of the dense layer units (neurons) are learned during the network training phase. For more detail about CNN architectures and training, see, for example, \citet{Indolina2018}. \par
\begin{figure}
    \centering
    \includegraphics[width=0.8\columnwidth]{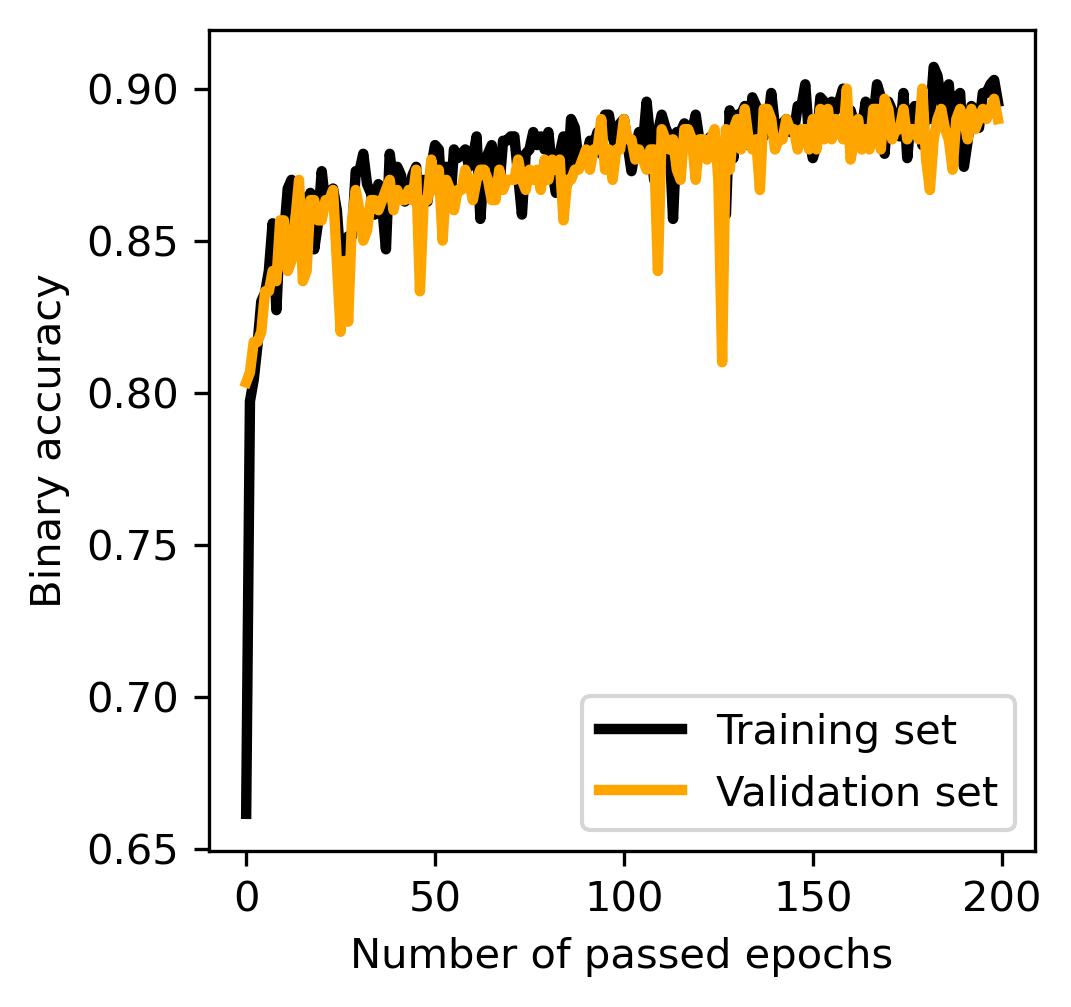}
    \caption{Evolution of the accuracy of the network predictions during the training process. The validation set accuracy (orange) closely follows the training set accuracy (black), showing the network does not overfit.}
    \label{figure:training_accuracy}
\end{figure}

 Our network has been trained for 200 epochs, using mini-batch training with a batch-size of 16. Training for more epochs does not significantly improve the training accuracy and leads to overfitting. For the network training, every network output with a value >0.5 is assigned to the label "binary". As the training progresses, the fraction of correctly labelled spectra increases. We show the improvement of the training and validation set accuracies during the 200 training epochs in Fig.~\ref{figure:training_accuracy}. The labelling accuracies for both sets increase together to a final value of $\sim$90\%. This close match between the two sets shows that our CNN does not overfit to the training data at any point during the training.

\begin{figure}
    \centering
    \includegraphics[width=0.8\columnwidth]{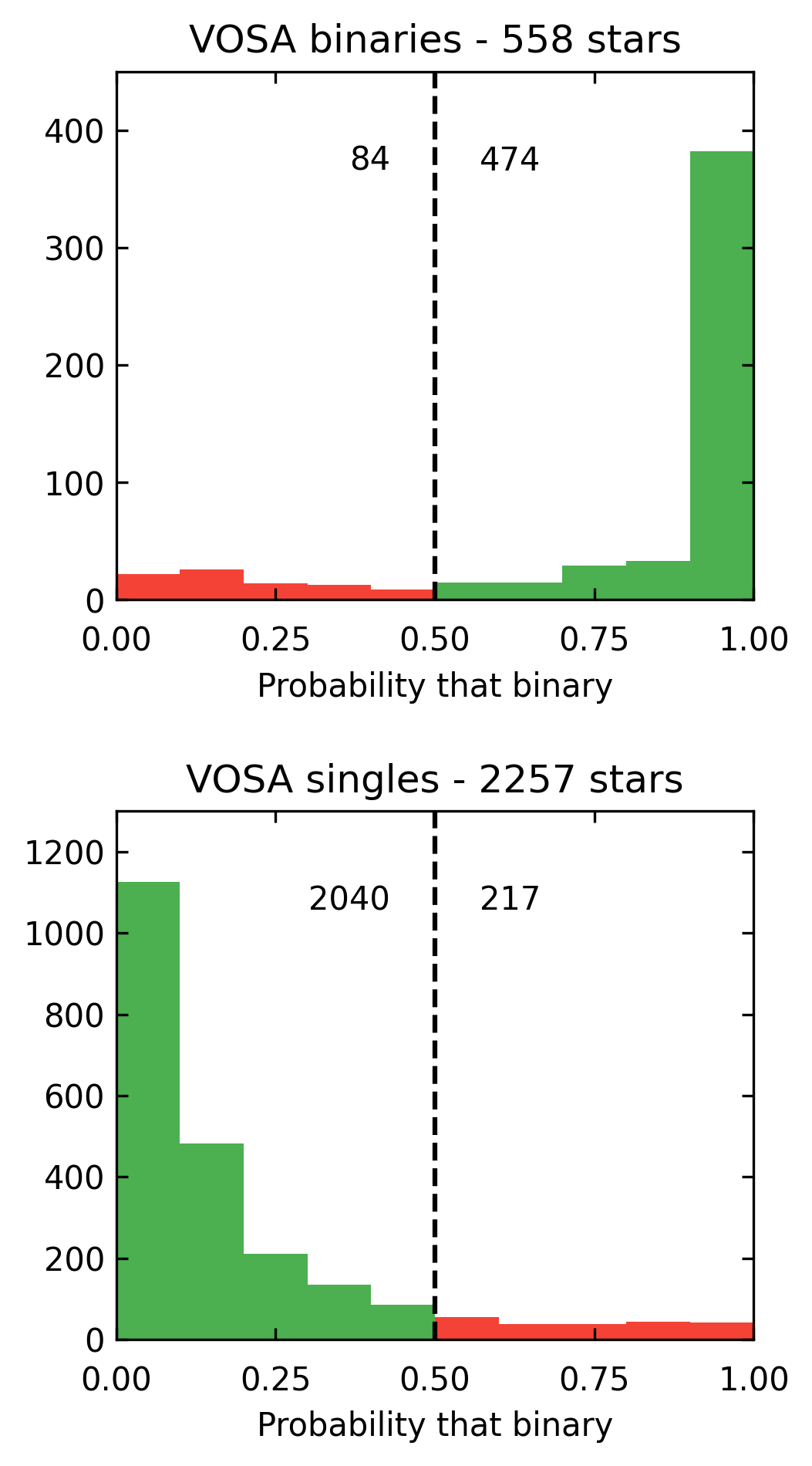}
    \caption{Distribution of CNN output probabilities for spectra to be "binary". Top panel: Distribution of probabilities for spectra that have been labelled binary by VOSA. Bottom panel: Same as top panel, but for VOSA single spectra. Numbers in the plot show how many spectra fall above or below the 50\% probability threshold. Green bars highlight agreement of CNN and VOSA labels, red bars show disagreement between the two methods.}
    \label{figure:netork_predictions}
\end{figure}

When we use our trained network to classify our full sample of 2815 spectra (of which 558 are labelled "binary" by VOSA), the accuracy decreases to $\sim$85\%. As can be seen from Fig.~\ref{figure:netork_predictions}, the reason for this drop in accuracy is that our network wrongly labels some VOSA single spectra as binary. \par

\begin{figure}
    \centering
    \includegraphics[width=0.8\columnwidth]{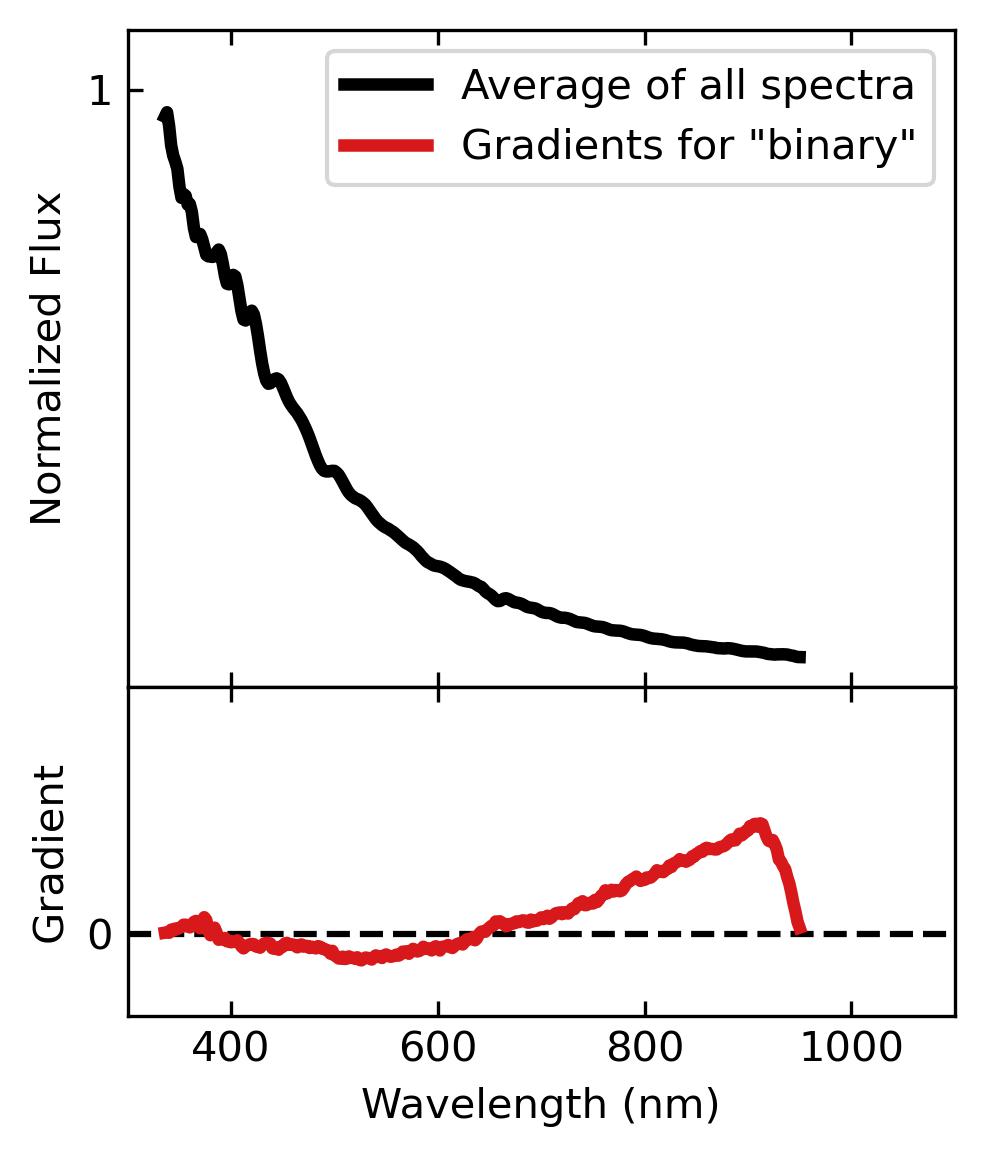}
    \caption{Top panel: Average of all GDR3~BP/RP spectra. Bottom panel: Network gradients for the prediction "binary" at every wavelength pixel.}
    \label{figure:network_gradients}
\end{figure}

Network gradients have been used to show that CNNs can label stellar spectra in a physically meaningful way (for example, \citealt{Ambrosch2023} and \citealt{Nepal2023}). These gradients describe how individual flux values in the input spectra influence the values of the CNN outputs. In our case, the network gradients show which parts of an input spectrum have the most influence on the probability for being binary. A positive gradient at a certain wavelength indicates that there is a positive correlation between flux value and output probability. The higher the flux at this wavelength, the higher the probability for being binary. Negative gradients indicate negative correlation between flux and output probability. Gradients are close to zero at parts of the spectrum that do not have an influence on the CNN output. Figure~\ref{figure:network_gradients} shows the gradients for the probability of being binary together with an average spectrum from our sample. At lower wavelengths, the gradient is close to zero. This means that in this part of the spectra, our CNN cannot effectively identify features that help it to decide whether a spectrum is binary or not. At wavelengths larger than 700~nm however, the gradient is positive and increases steadily. We can therefore conclude that our network assigns binary probabilities based on the infrared flux in our GDR3 BP/BR spectra. This is consistent with how VOSA identifies binaries based on flux excess towards redder bands.

\section{Evaluation and comparison of the prediction metrics for the different classification methods.}
\label{sec:comparison}

 In our analysis, we adopted several key metrics to evaluate the performance of our classification models \citep[see e.g.][]{Stehman1997}. This has been done considering the VOSA classification as the true one. Thus, TP are the true positives, which occur when the model accurately predicts a binary while FP are the false positives, that is, when a binary is predicted incorrectly; on the contrary, the TN are the true negatives, which occur when a single is correctly predicted, while the false negatives (FN) occur when a single is predicted incorrectly. These metrics include True Positive Rate (TPR) (Eq.~\ref{eq:TPR}), which measures the proportion of binary stars correctly identified as such, and True Negative Rate (TNR) (Eq.~\ref{eq:TNR}), which measures the proportion of single stars correctly identified. Additionally, we assessed Accuracy (Eq.~\ref{eq:Accuracy}), which represents the overall proportion of correctly classified instances, considering both binary and single stars. The Balanced Accuracy (Eq.~\ref{eq:BalancedAccuracy}) accounts for the imbalance between the two classes by taking the average of TPR and TNR. Precision Predictive Value (PPV) (Eq.~\ref{eq:PPV}) focuses on the proportion of correctly predicted binary stars among all predicted binary stars, while Negative Predictive Value (NPV) (Eq.~\ref{eq:NPV}) focuses on the proportion of correctly predicted single stars among all predicted single stars. Lastly, the F1 Score (Eq.~\ref{eq:F1ScoreP} and \ref{eq:F1ScoreN}) provides a balance between precision and recall, considering both the proportion of correctly predicted binary stars and the ability to capture all binary and single stars.

\begin{equation}
\label{eq:TPR}
\quad TPR = \frac{TP}{TP + FN}
\end{equation}

\begin{equation}
\label{eq:TNR}
\quad TNR = \frac{TN}{TN + FP}
\end{equation}

\begin{equation}
\label{eq:Accuracy}
\quad Acc = \frac{TP + TN}{TP + TN + FP + FN}
\end{equation}

\begin{equation}
\label{eq:BalancedAccuracy}
\quad Acc \, (bal) = \frac{1}{2} \times \left( TPR + TNR \right)
\end{equation}

\begin{equation}
\label{eq:PPV}
\quad PPV = \frac{TP}{TP + FP}
\end{equation}

\begin{equation}
\label{eq:NPV}
\quad NPV = \frac{TN}{TN + FN}
\end{equation}

\begin{equation}
\label{eq:F1ScoreP}
\quad F1 \,  score \, (P) = 2 \times \frac{PPV \times TPR}{PPV + TPR}
\end{equation}

\begin{equation}
\label{eq:F1ScoreN}
\quad F1 \,  score \, (N) = 2 \times \frac{NPV \times TPR}{NPV + TPR}
\end{equation}

\begin{figure}
  \resizebox{\hsize}{!}{\includegraphics{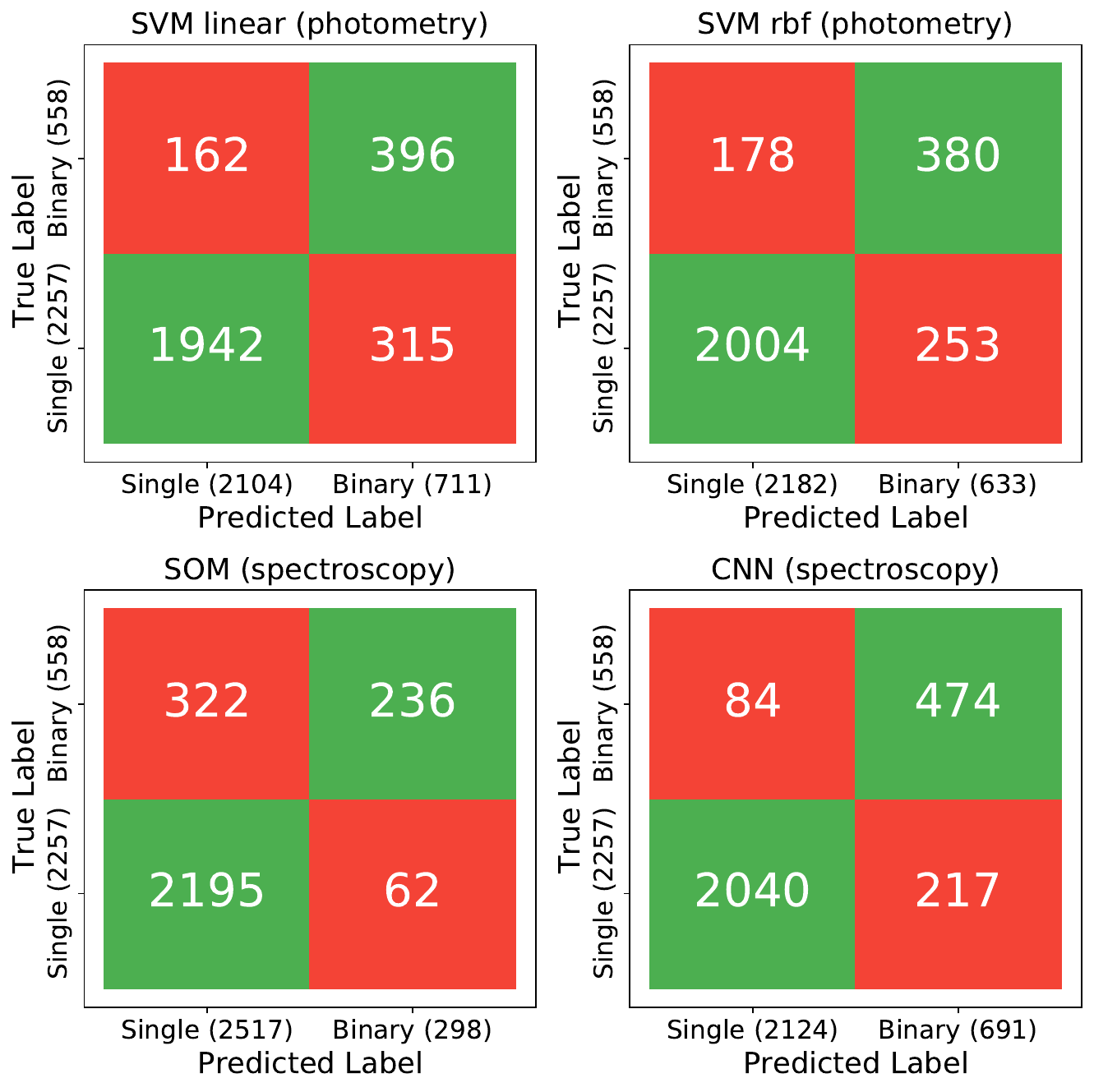}}
  \caption{Confusion matrix for the different classification methods. Each matrix displays the performance of a method in distinguishing between 'Single' and 'Binary' classes. The cells shaded in green represent true positive and true negative predictions, indicating correct classifications, while cells shaded in red represent false positive and false negative predictions, indicating incorrect classifications.}
  \label{fig:confusion_matrix}
\end{figure}

  In 
  %the 
  Fig.~\ref{fig:confusion_matrix}, we present the confusion matrix, showcasing the true positives (binaries) and negatives (singles), along with the false positives and negatives, resulting from the predictions made by the four methods relative to the classification provided by VOSA. In Figs.~A.1 and A.2 we compare the labels  obtained with the 5 methods (including VOSA) with each other as a colour-coded matrix, where the number represents the common stars with the same label (true positives and negatives). For instance, based on these figures, it's evident that CNNs and linear w-SVMs are the most effective in identifying binaries (543 objects classified as binaries in common). In Tab.~\ref{tab:prediction_metrics} we show the prediction quality metrics, containing the percentages of true labels with respect to those obtained individually with VOSA.  These results represent the performance of each method's classifications and how well they align with those of VOSA in classifying stars as binaries or singles based on photometry (SVM) and spectroscopy (SOM and CNN). As we can see, SVM (linear) and SVM (rbf) achieve similar True Positive Rates (TPR) around 68-70\%, indicating their ability to correctly identify binary stars using photometric data. SOM shows a lower TPR of 42\%, suggesting it may struggle more with binary classification based on spectroscopic data. In Fig.~A.3, we present the classification results of SOMs and CNNs on a CMD, thus combining spectroscopy with photometry. As evident, single stars predominantly occupy the bluest region, while binaries are predominantly found in the redder region, as expected. It is notable that CNNs exhibit a superior capability in classifying binaries compared to SOMs, which appears more conservative. 
  CNN demonstrates the highest TPR, 85\%, indicating strong binary classification capabilities using spectroscopic information. True Negative Rates (TNR) are consistently high across all methods, ranging from 86\% to 97\%, indicating their proficiency in identifying single stars based on both photometric and spectroscopic data. Overall Accuracy rates vary between 83\% and 89\%, with CNN showing the highest accuracy, suggesting its effectiveness in both binary and single star classification using spectroscopic data. However, our sample of singles/binaries is unbalanced at approximately 80/20\%. The balanced accuracy considers the balance between TPR and TNR, that is, the contribution of both classes equally (50/50\%), giving the same weight to the accuracy in the classification of both classes. We got values ranging from 78\% to 87\%, reflecting the overall performance in both binary and single star classification using photometric and spectroscopic data. Positive Predictive Value (PPV) ranges from 56\% to 79\%, indicating the proportion of correctly identified binary stars among all stars classified as binaries using photometric and spectroscopic data. Negative Predictive Value (NPV) ranges from 87\% to 96\%, showing the proportion of correctly identified single stars among all stars classified as singles using photometric and spectroscopic data. F1 Score (P), which considers both precision and recall, varies from 55\% to 75\%, providing a combined measure of binary classification performance using photometric and spectroscopic data. F1 Score (N) indicate strong performance across all methods in correctly classifying the negative (binary) instances, with values ranging from approximately 88\% to 93\%. All these metrics together offer insights into the effectiveness of each method in distinguishing between binary and single stars, with CNN exhibiting the most promising performance across multiple evaluation criteria when utilizing spectroscopic data.
  
  %\textcolor{magenta}{Question:Why do all classifiers perform better for singles than for binaries? Might the 80/20 balance be playing a role here? }
 
 We emphasize that to classify a star as single or binary in the methods described in this work, we have considered that the probability of belonging to one class is greater than 50\%. However, to classify a star as single or binary robustly we recommend only considering those stars that agree on the classification by all methods or with a high probability in several of them. Upon analysing 2815 stars, we identified 260 objects that meet the criteria for binary classification based on all four of our photometric and spectroscopic methods, while 1947 are classified as single stars.  This shows a strong concordance rate of 78\% across all four classification techniques (linear and RBF SVMs based in photometry, as well as spectroscopy using SOM and CNNs), with a 22\% discrepancy in classifications. This criterion of agreement in the classification offers statistical reliability, diminishes classification errors, ensures consistency among methods, and enhances the trustworthiness of result interpretation. In the Appendix we provide an extract from the online table (Table~\ref{tab:labels}) with the predicted label by the different methods and probability of each object being binary. In the same table we also provide a column with a quality flag with values between 0-5, where a value of 0 means that all 5 methods classify the object as single and a value of 5 means that all 5 methods classify it as binary. It is worth mentioning again that each method tested has a different approach to the problem, so the SOM is an unsupervised method that does not know the labels "a priori", while SVM and CNN are supervised methods. While for the SVMs we have used photometry, for the other methods we have used spectroscopy.

\begin{table}[h]
    \centering
    \caption{Prediction quality metrics for different methods with respect to VOSA.}
    \begin{tabular}{lcccc}
        \hline\hline
        Metric & SVM (l) & SVM (rbf) & SOM & CNN \\
        \hline
        TPR  & 70.97\% & 68.10\% & 42.29\% & 84.95\%\\
        TNR  & 86.04\% & 88.79\% & 97.25\% & 90.34\%\\
        Accuracy  & 83.05\% & 84.69\% & 86.35\% & 89.31\%\\
        Accuracy (bal.) & 78.50\% & 78.44\% & 69.78\% & 87.66\%\\
        PPV  & 55.70\% & 60.03\% & 79.19\%& 68.59\%\\
        NPV  & 92.30\% & 91.84\% & 87.21\% & 96.04\% \\
        F1 Score (P)  & 62.41\% & 63.81\% & 55.14\% & 75.89\%\\
        F1 Score (N) & 88.08\% & 90.28\% & 91.01\% & 93.08\%\\
       \hline
    \end{tabular}
    \label{tab:prediction_metrics}
\end{table}

 Finally, a statistical comparison was conducted among the probabilities of being a binary star across the spectroscopic methods. In Fig.~A.4 %\ref{Appendix}, 
 we show the cumulative distribution function (CDFs) and the results of the Kolmogorov-Smirnov (K-S) statistic test \citep{kolmogorov33, smirnov1939estimate} applied to the probability of being binary and computed using \texttt{scipy.stats} \citep{Virtanen2020}. The K-S test provides an objective measure of how similar or different these probability distributions are and allows evaluating the consistency between results obtained by both classification methods. Additionally, it can help to identify regions within the distributions where significant discrepancies exist. We obtained a good agreement between the probability distribution using both methods, with very low p-values (< 0.05). This means that there is a very low probability that those results occurred by random chance, and we have stronger evidence to reject the null hypothesis. We also found that the maximum absolute difference between the cumulative distributions of probabilities of being binary is at about P = 0.5. As we can see in Fig.~A.4, the SOMs reach a probability of being binary > 0.5 at a higher density than the CNNs, which indicates that the latter classify more stars as binary than the former, which is more conservative in its predictions.  
 
 %\textcolor{teal}{} It is worth highlighting the probability distribution obtained with linear SVM and the SOM, which is very similar (see Fig. $\ref{fig:KS_SVM_linear_vs_SOM}$).

%\section{\textcolor{teal}{Exploring uncommon objects in large sets of \textit{Gaia} DR3 spectra.}}
%\label{sec:88}
\section{Possible unveiling of outliers in large \textit{Gaia} DR3 spectra sets.}
\label{sec:88}

In this section, we explore further ways to analyse and categorize larger sets of spectra. When applying machine learning techniques to unlabelled spectra, it is important to understand the structure and internal properties of the full set. This allows us, for example, to identify outlier spectra or other special groups of spectra that may need to be treated separately. This type of pre-analysis enables us to interpret the outputs of our various ML techniques, specially when dealing with very large data sets. We tested this idea with a selected and well-known hot subdwarf sample with the objective of laying groundwork for future steps, as we aim at further analysing (in preparation, in a separate work) the $\sim$61,000 candidates identified by \citet{Culpan2022}, by means of their \textit{Gaia} spectra, enabling us to initially distinguish objects that do not fit the hot sds profile.

\subsection{The 'cosine similarity' of single and binary spectra of a hot sds 'golden sample'.}
\begin{figure}
  \resizebox{\hsize}{!}{\includegraphics{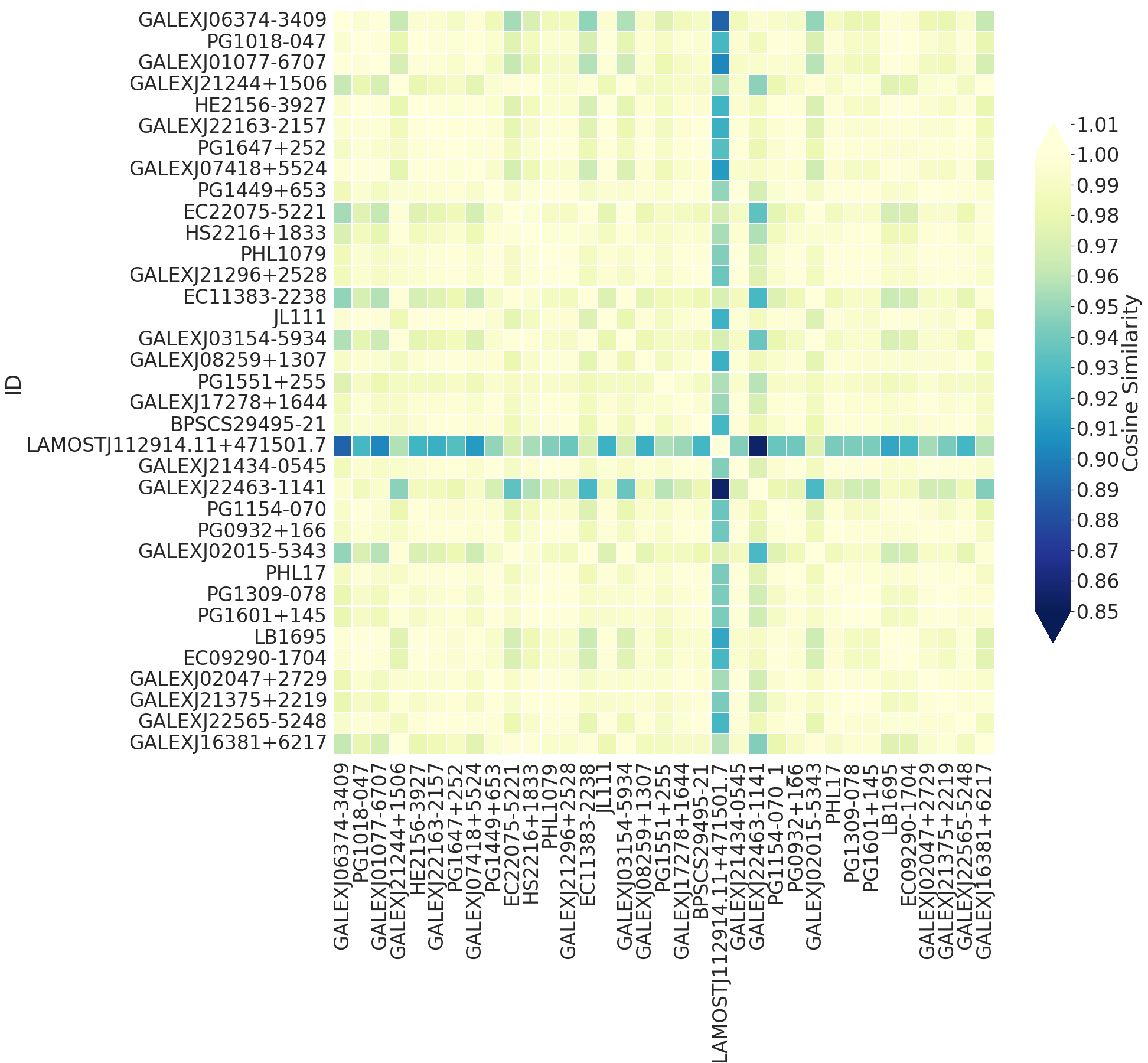}}
  \caption{Heatmap of 35 binary stars from a chosen subsample from \citet{Solano22} and colour-coded according to their cosine similarity.}
  \label{fig:35_cosine}
\end{figure}

\begin{figure}
  \resizebox{\hsize}{!}{\includegraphics{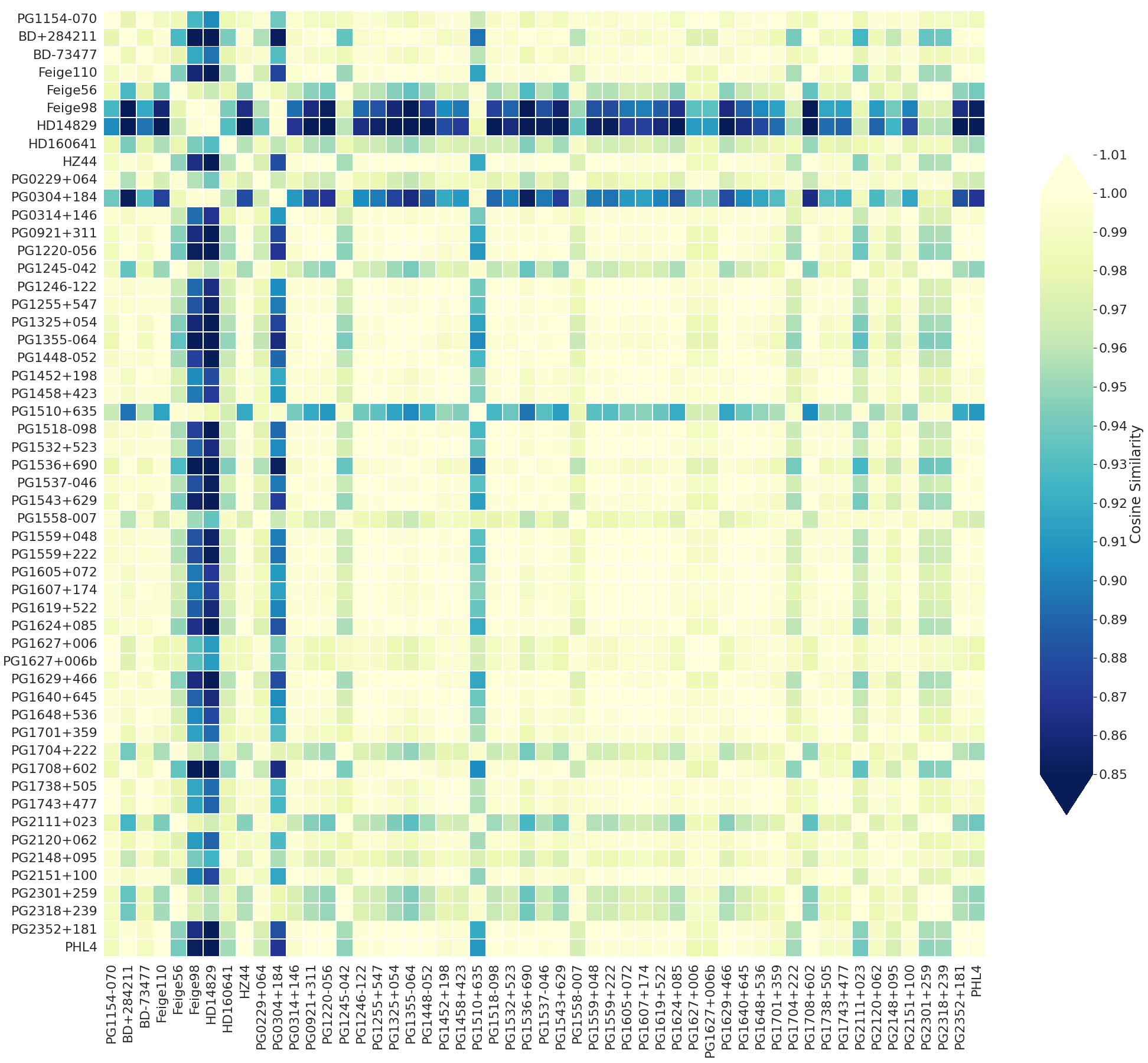}}
  \caption{Heatmap of 53 single stars from a chosen subsample from \citet{Drilling13} and colour-coded according to their cosine similarity.}
  \label{fig:53_cosine}
\end{figure}

In Sec.~\ref{sec:spectroscopy} we made a pre-analysis of the 2815 GDR3 BP/RP spectra, finding a small percentage of them as anomalous. In order to investigate these spectra in more detail, we chose a so-called 'golden sample' of 88 targets, to which we applied the same technique. These well-defined hot sds are 35 binaries (from \citet{Solano22}) and 53 singles (from \citet{Drilling13}), which were selected as follows. 
%After a careful final data inspection, \citet{Solano22} retained 42 targets with sufficiently good SEDs for two-body fitting attempts and for effective temperature determinations. After a cross-match with GRD3, we found 35 of them to display BP/RP spectra. On the other hand, \citet{Drilling13} provided an MK (Morgan-Keenan)-like classification system which is still used as a reference for single hot sd spectral classification: 53 objects out of their Tab. 1 were also found to display \textit{Gaia} DR3 BP/RP spectra. For these selected hot sds, we compared the 88 normalized \textit{Gaia} DR3 BP/RP spectra with each other using the 'cosine similarity' measure implemented using the {\sc scikit-learn} package \citep{scikit-learn11}. 
\citet{Drilling13} provided an MK (Morgan-Keenan)-like classification system which is still used as a reference for single hot sd spectral classification. Binary contamination was excluded from their study and the targets retained by the authors are considered a representative single hot subdwarfs and blue horizontal-branch stars sample. 53 objects out of their Tab. 1 were found to have \textit{Gaia} DR3 BP/RP spectra and, therefore, suitable for our analysis. For the case of binary hot sds classification, we are not aware of a similar published system to the one provided by \citet{Drilling13} for single objects. Therefore, we selected 35 binary hot sds from the work by \citet{Solano22}, with \textit{Gaia} BP/RP spectra. That is to say, that, after a careful final data inspection, \citet{Solano22} retained 42 targets with sufficiently good SEDs for two-body fitting attempts and for effective temperature determinations. Out of a cross-match of those 42 binaries with GDR3, we retained the above-mentioned 35 of them to have BP/RP spectra.
For these selected hot sds, 53 singles from \citet{Drilling13} and 35 binaries from \citet{Solano22}, we compared the 88 normalized \textit{Gaia} DR3 BP/RP spectra with each other using the 'cosine similarity' measure implemented using the {\sc scikit-learn} package \citep{scikit-learn11}.
To do this, we considered these 88 normalized spectra as vectors and derived the 'cosine similarity' by using Eq.~\ref{eq:cosine_similarity} as defined in %section 
Sec.~\ref{sec:spectroscopy}. The results are shown in a colour-coded matrix in Fig.~\ref{fig:35_cosine} for binaries and in Fig.~\ref{fig:53_cosine} for singles. This 'similarity matrix' provides a visual representation of the degree of similarity between each pair of spectra, highlighting patterns and relationships within the dataset. As can be seen, one star (LAMOSTJ112914.11+471501.7) stands out especially among the binaries and 4 stars (HD14829, Feige98, PG0304+184 and PG1510+635) among the singles, with a few more in between. For these, the mean value of the 'cosine similarity' is significantly lower than the others ($\leq$ 0.95), which indicates that the angle of the vectors that represent their spectra is greater and therefore they differ more from the others. LAMOSTJ112914.11+471501.7 had been classified by the SVMs and CNNs in the preceding sections as binary, which is in agreement with VOSA. \textit{Gaia}'s spectra are reconstructed from a series of basis functions, and these deviations (which are not real) occasionally emerge, particularly in the spectra of the five stars under discussion. Thus, the cosine similarity method proves effective in detecting outliers or stars significantly different from the rest.

%\sout{The spectra of the star HD14829 is the most different and contrasts the most with the others, with a mean 'cosine similarity' of 0.89. Applying the classification method described in subsection \ref{subsec:color-mag} based on the colour-magnitude diagram, the star "LAMOSTJ112914.11+471501.7" would have a high probability of being single and therefore it would have been misclassified}.

%For a better visualization of the differences, we have reduced the dimensionality of the similarity matrix using Principal component analysis (PCA) through the Singular Value Decomposition with the {\sc scikit-learn} package \citep{scikit-learn11}. Then we have represented it with a network graph (Fig. \ref{fig:88_redes_pca} in the Appendix) using {\sc NetworkX} package \citep{Aric_11} in which each node represents a spectrum and each edge the degree of similarity. So that the closer nodes have spectra that are more similar to each other and the farther nodes are more different. As can be seen in the aforementioned Fig. \ref{fig:88_redes_pca}, the furthest nodes correspond to the spectra of the stars we mentioned before. Thus, the star HD14829 is located at the farthest vertex.

After this, the next step is to inspect the spectra of those more differentiated stars, which is shown in Figs.~A.5 to A.9. As we can see, those 5 stars fit into a common category, with characteristics that differentiate them. This is especially seen in the features at shorter wavelengths ($\lambda$ < $\sim$ 400 nm) and because for larger intervals of the spectra, the flux is higher than the others. This leads us to classify the spectra into different categories according to their common features in their spectra.

\begin{figure}
  \resizebox{\hsize}{!}{\includegraphics{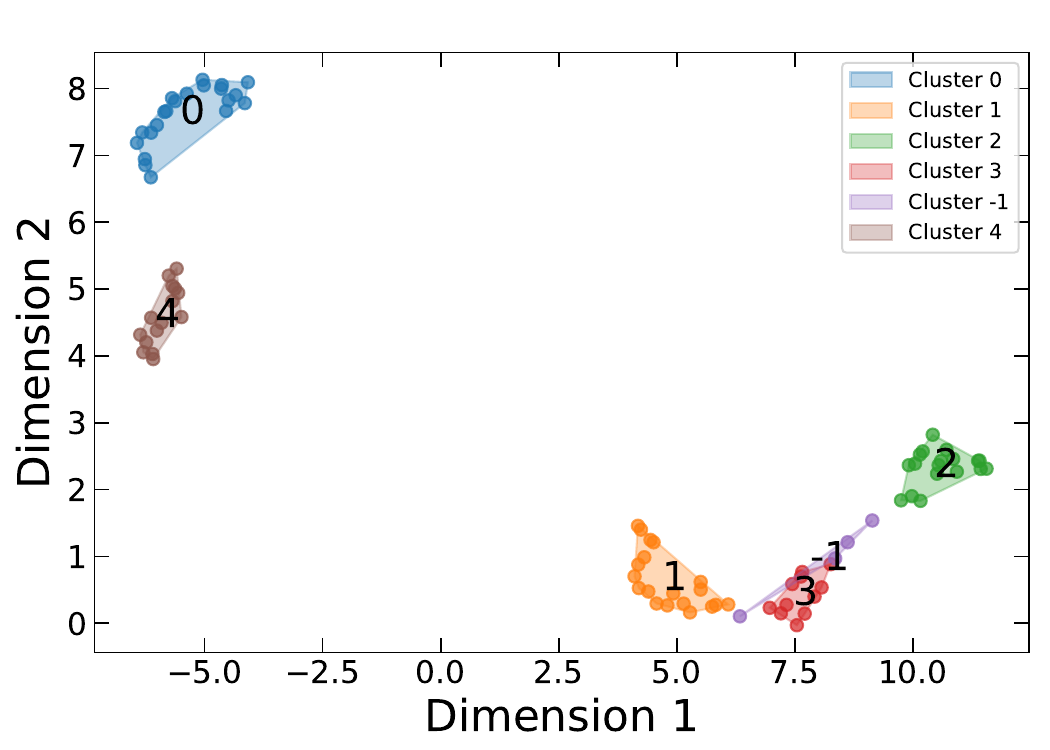}}
  \caption{Visualization of the similarity matrix reduced to 2 dimensions using UMAP. Groups of spectra that follow similar patterns are clustered using DBSCAN.}
  \label{fig:umap+dbscan}
\end{figure}

Thus, in search of patterns within the spectra, we also used the UMAP technique to reduce the dimensionality of the similarity matrix (from 88x88 to a two-dimensional space) and then applied the Density-Based Spatial Clustering of Applications with Noise (DBSCAN) algorithm through the {\sc scikit-learn} package \citep{scikit-learn11} to perform clustering on the reduced data. This method is able of differentiating the spectra into several groups, which is shown in Fig.~\ref{fig:umap+dbscan}. As we can see, clusters 0 and 4, which are closest in the aforementioned Fig.~\ref{fig:umap+dbscan}, are characterized by having a decreasing spectrum at increasing wavelength, occupying the lowest fluxes compared to the others (Figs.~A.10 to A.15). Then we have clusters 1, 2 and 3, with higher fluxes than clusters 0 and 4 at longer wavelengths, and more pronounced features below at around 400-450 nm. Cluster -1 would correspond to "outlier" or noise spectra, since they do not meet the criteria to be part of a cluster and they are difficult to classify. Finally, the 5 stars that we mentioned above with peculiar patterns would be included in cluster 2, which are characterized by having the highest fluxes, and in most of the cases with the special feature at short wavelengths ($\lambda$ < $\sim$ 400 nm). 

When confronting the clusters with the detailed spectral types from \citet{Drilling13}, we observe distinct patterns in the distribution of spectral subtypes. Clusters 4 and 0 are dominated by early subtypes of sdO and sdB, as well as sdBN types. Clusters 1 and 3 features intermediate subtypes of sdB. Finally, cluster 2, with the highest mean fluxes, includes a variety of later subtypes of sdB and features sdA and sdOC types. Clusters with lower mean fluxes (4 and 0) contain more early subtypes, whereas clusters with higher mean fluxes (1, 3 and especially 2) include more late subtypes and varied types, according to \citet{Drilling13} classification. The "outliers" are identified as blue horizontal branch (BHB) stars, distinguishable by their prominent Balmer lines that can be clearly detected in the low-resolution BP/RP spectra, as shown in the Appendix figures.

%\sout{In Figs. \ref{fig:cluster0} to \ref{fig:cluster2} in the Appendix we show the different groups of spectra. Identifying the stars of each cluster, for example, we can observe that in general the clusters to the left of the Fig. \ref{fig:umap+dbscan} correspond to the objects classified as singles by \citet{Drilling13}, while those located to the right correspond to the \citet{Solano22} binary objects. 

%However, this method does not seem conclusive for classifying stars into singles or binaries in this reduced sample.

\subsection{The relative difference between singles and binaries.}
\label{sec:Relative difference between singles and binaries}
\begin{figure}
  \resizebox{\hsize}{!}{\includegraphics{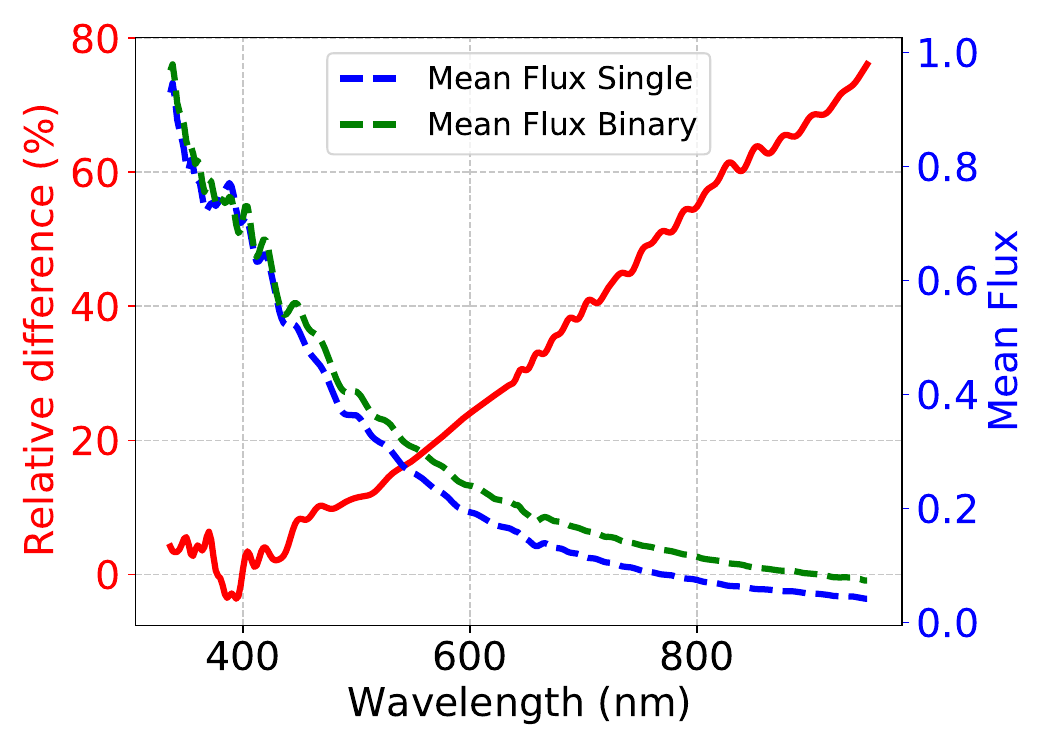}}
  \caption{Relative difference and mean fluxes between single and binary stars for our subsample of 53 singles and 35 binaries.}
  \label{fig:88_dif}
\end{figure}

On the other side, we conducted an analysis of spectral flux data for the 88 stars differentiating between single and binary stars. After processing the data, we computed the mean flux for both types and calculated the relative difference. Visualization was achieved through a dual-axis plot, with the left axis displaying relative differences in percentages and the right axis showing mean flux values (see Fig.~\ref{fig:88_dif}). Visualizing this relative difference across the spectrum helps identify specific wavelengths where the two types of stars exhibit significant distinctions in their average flux values. The results show that the two classes become more differentiated as the wavelength increases, reaching relative differences of 60-80\% beyond 800 nm. This is in agreement with what we saw in 
%section 
Sec.~\ref{sec:CNN} for the whole sample and is consistent with \citet{Solano22}'s conclusions using VOSA tools. 
The increasing trend in relative difference, coupled with a relatively constant difference between mean flux values of single and binary stars, may suggest heightened sensitivity or distinct spectral features at specific wavelengths. This indicates that certain wavelengths play a crucial role in distinguishing between single and binary stars, possibly due to pronounced spectral characteristics, the presence of specific chemical elements, or unique physical processes.Basically, what we are noticing here is the contribution of the secondary to the net flux. If the secondary is cooler, it will mostly emit at redder wavelengths, making this range quite appropriate to detect its contribution. The different nature of the companion on binary systems with hot sds would support this. The observed trends underscore the potential utility of certain wavelengths for discerning between binary and single stars based on their spectral features. Further exploration of these distinct regions can contribute to a more detailed understanding of the distinguishing factors between these stellar types.

\section{Summary and conclusions}\
\label{sec:conclusions}
In this study, we investigate the classification of single/binary hot subdwarf sources using various supervised and unsupervised classification criteria along with different \textit{Gaia} DR3 datasets (both photometry and BP/RP spectroscopy). Our goal is to assess the extent to which we can replicate the classification provided in \citet{Solano22} using VOSA and visual inspection of Spectral Energy Distributions. 
%(SEDs).

Our findings reveal that, in general, all methods used demonstrates a high level of agreement with VOSA's manual classification, achieving a reproducibility rate of near $\sim$70-90\% for all cases. This suggests that SVM, SOM and CNN techniques can effectively classify single/binary bona-fide sources with accuracy comparable to human inspection using VOSA tools. The method based on spectroscopy using the CNN technique deserves special attention, which reaches 84.94\% reproducibility for the detection of binaries. In global terms, the CNN reaches the best accuracy with near $\sim$90\% of successes for the unbalanced and balanced samples. Techniques based on photometry using SVMs reach a $\sim$80\% of accuracy. We also found an agreement of 78\% in the classification using the four methods (through photometry and spectroscopy). The single/binary classification ratios achieved by the different methods are as follows: VOSA (80/20), linear w-SVM (75/25), RBF w-SVM (77/22), SOM (89/11) and CNN (75/25). Other techniques such as the 'cosine similarity' and UMAP have also been applied, for pre-analysis of the 2815 sources with \textit{Gaia} DR3 spectra and, to gain further insight, to a smaller and well-defined (88) subsample, proving very effective for the detection of outliers and peculiar spectra. 

For future studies, we plan to leverage the \citet{Culpan2022} sample, which provides improved accuracy and a larger dataset compared to \citet{Geier19}. The \citet{Culpan22} catalogue addresses issues with \textit{Gaia} astrometry, particularly for close-by stars, thereby reducing the risk of misclassifying A-type MS stars as composite sdB binary candidates, as highlighted by \citet{Dawson2024}. Taking advantage of this, our next step (in preparation) is to tackle the $\sim$61,000 candidates in \citet{Culpan2022} to identify objects whose spectra significantly differ from those expected of singles or binaries. In extensive datasets, such as those in the aforementioned \citet{Culpan2022}, we anticipate significant contamination with DA white dwarfs, which also exhibit strong Balmer lines. Accurate identification of these DA white dwarfs is crucial. This pre-analysis aims to create a more refined subset of hot subdwarf candidates, which can then be analysed further using their GDR3 BP/RP spectra and the classification methods described in this paper. The methods above can also prove useful for finding misclassified objects. All techniques are complementary to each other and allow the comparison/validation of results using different sources, whether photometry or spectroscopy. \\

Furthermore, our analysis highlights the challenges associated with classifying a large sample of sources where the composition is uncertain. While the classification accuracy is promising, caution is warranted when dealing with potentially contaminated samples. Future research could focus on refining classification techniques to mitigate the effects of contamination and improve the reliability of automated classification methods. Overall, this study contributes to the ongoing efforts in automating the classification of astronomical sources like single/binary hot sds, providing insights into the effectiveness of different methodologies and their applicability in large-scale surveys. Further investigations are able to explore additional techniques and validate the robustness of the classification results, especially in the presence of complex data sets and ambiguous source compositions.

\section*{Data availability}
Figures A.1 - A.15 are available in the Zenodo platform at \href{https://doi.org/10.5281/zenodo.13841865}{https://doi.org/10.5281/zenodo.13841865}.

\begin{acknowledgements}
We sincerely thank the anonymous referee for her/his valuable guidelines and insightful comments, which have significantly enhanced the quality of this work. This research has made
use of the Spanish Virtual Observatory (\href{https://svo.cab.inta-csic.es}{https://svo.cab.inta-csic.es}) project funded by
MCIN/AEI/10.13039/501100011033/ through grant PID2020-112949GB-I00. Also made use of GUASOM \citep{Fustes14,Alvarez22}, Scikit-learn Machine Learning \citep{scikit-learn11}, NetworkX \citep{Aric_11}, Seaborn \citep{Waskom2021}, TopCat \citep{Taylor2005}, Pandas \citep{pandas2020} and Matplotlib \citep{Hunter2007}. This research has made extensive use of NASA's Astrophysics Data System Bibliographic Services. This work has made use of data from the European Space Agency (ESA) \textit{Gaia} mission, processed by the \textit{Gaia} Data Processing and Analysis Consortium (DPAC). Funding for the DPAC has been provided by national institutions, in particular, the institutions participating in the \textit{Gaia} Multilateral Agreement. This research has made use of the Simbad database and the Aladin sky atlas, operated at CDS, Strasbourg, France. The authors have also made use of the VOSA software, developed under the Spanish Virtual Observatory project supported by the Spanish MINECO through grant PID2020-112949GB-I00. Funding from Spanish Ministry project PID2021-122842OB-C22, Xunta de Galicia ED431B 2021/36 and PDC2021-121059-C22 is acknowledged by the authors. This work was funded by the Spanish MCIN / AEI / 10.13039 / 501100011033 and European Union Next Generation EU/PRTR through grant PID2021-122842OB-C22 and the Horizon Europe [HORIZON-CL4-2023-SPACE-01-71], SPACIOUS project funded under Grant Agreement no. 101135205. CVV and AU thank the MW-\textit{Gaia} COST Action "Revealing the Milky Way with \textit{Gaia}" CA18104 for its support through a Short-term scientific mission (STSM) at the University of Vigo and to Erasmus+Staff for supporting a scientific visit of CVV to the aforementioned university. MAA, MM, RSG and JCD also acknowledge support from CIGUS CITIC, funded by Xunta de Galicia and the European Union (FEDER Galicia 2021-2027 Program) through grant ED431G 2023/01. This work is in the memory of Carlos Rodrigo (†), deceased during the preparation of this work.

\end{acknowledgements}

\bibliographystyle{aa} % style aa.bst
\bibliography{Bibliography}

\begin{appendix}
\section{Complementary material}
\label{Appendix}

Figures A.1 - A.15 are available in the Zenodo platform at \href{https://doi.org/10.5281/zenodo.13841865}{https://doi.org/10.5281/zenodo.13841865}.

\begin{table*}
\caption{Classification labels for the different methods used in this work, where 0 means single and 1 binary. The probabilities of being binary and a flag that indicates the agreement between the different methods are also provided.}
\label{tab:labels}
\scalebox{0.97}{
\begin{tabular}{lcccccccc}

\hline\hline
object& VOSA& w-SVM$_{linear}$& w-SVM$_{RBF}$& SOM& CNN& P$_{binary}$ SOM& P$_{binary}$ CNN & flag\\
\hline
  PG0255+029 & 0 & 0 & 0 & 0 & 0 & 0.083 & 0.155&0\\
  PG0240+046 & 0 & 0 & 0 & 0 & 0 & 0.038 & 0.340&0\\
  PG0322+078 & 0 & 1 & 1 & 0 & 1 & 0.179 & 0.564&3\\
  LAMOSTJ032133.19+081131.6 & 0 & 1 & 0 & 1 & 0 & 0.778 & 0.297&2\\
  LAMOSTJ034208.81+090220.7 & 0 & 1 & 1 & 0 & 1 & 0.391 & 0.766&3\\
\hline
\end{tabular}}
\tablefoot{Only the first five rows of this table are displayed here. A machine-readable version of the complete table  is accessible at the CDS.}
\end{table*}

\end{appendix}


\end{document}